\documentclass[aps,pra,reprint,superscriptaddress,showpacs,showkeys,floatfix]{revtex4-1}
\usepackage{graphicx}
\usepackage{lipsum}
\usepackage{color}
\usepackage{xcolor}
\usepackage{amssymb}
\usepackage{amsmath}
\usepackage{epsfig}
\usepackage{bm}
\usepackage[american]{babel}
\usepackage{hyperref}
\usepackage{braket}
\usepackage{tabularx}
\usepackage{wasysym}
\usepackage[T1]{fontenc} 
\usepackage{txfonts} 
\usepackage{comment}
 
\DeclareGraphicsExtensions{.pdf,.png,.jpg}
\hypersetup{colorlinks=true,urlcolor=blue,breaklinks=true,citecolor=blue,linkcolor=blue,pdfstartview=FitH,pdfpagemode=UseNone}

\begin{document}

\title{Fidelity of time-bin entangled multi-photon states from a quantum emitter}

\date{\today}

\author{Konstantin Tiurev} \email{konstantin.tiurev@gmail.com}
\affiliation{Center for Hybrid Quantum Networks (Hy-Q), The Niels Bohr Institute, University~of~Copenhagen,  DK-2100  Copenhagen~{\O}, Denmark}

\author{Pol~Llopart~Mirambell}
\affiliation{Center for Hybrid Quantum Networks (Hy-Q), The Niels Bohr Institute, University~of~Copenhagen,  DK-2100  Copenhagen~{\O}, Denmark}

\author{Mikkel~Bloch~Lauritzen}
\affiliation{Center for Hybrid Quantum Networks (Hy-Q), The Niels Bohr Institute, University~of~Copenhagen,  DK-2100  Copenhagen~{\O}, Denmark}

\author{Martin~Hayhurst~Appel}
\affiliation{Center for Hybrid Quantum Networks (Hy-Q), The Niels Bohr Institute, University~of~Copenhagen,  DK-2100  Copenhagen~{\O}, Denmark}

\author{Alexey~Tiranov}
\affiliation{Center for Hybrid Quantum Networks (Hy-Q), The Niels Bohr Institute, University~of~Copenhagen,  DK-2100  Copenhagen~{\O}, Denmark}

\author{Peter~Lodahl}
\affiliation{Center for Hybrid Quantum Networks (Hy-Q), The Niels Bohr Institute, University~of~Copenhagen,  DK-2100  Copenhagen~{\O}, Denmark}

\author{Anders~S{\o}ndberg~S{\o}rensen}
\affiliation{Center for Hybrid Quantum Networks (Hy-Q), The Niels Bohr Institute, University~of~Copenhagen,  DK-2100  Copenhagen~{\O}, Denmark}

\begin{abstract}
We devise a mathematical framework for assessing the fidelity of multi-photon entangled states generated by a single solid-state quantum emitter, such as a quantum dot or a nitrogen-vacancy center. Within this formalism, we theoretically study the role of imperfections present in real systems on the generation of time-bin encoded Greenberger--Horne--Zeilinger and one-dimensional cluster states. We consider both fundamental limitations, such as the effect of phonon-induced dephasing, interaction with the nuclear spin bath, and second-order emissions, as well as
technological imperfections, such as branching effects, non-perfect filtering, and photon losses. 
In a companion paper, we consider a particular physical implementation based on a quantum dot emitter embedded in a photonic crystal waveguide and apply our theoretical formalism to assess the fidelities achievable with current technologies.
\end{abstract}

\maketitle

\section{Introduction}\label{section:intro}
A  reliable  source  of  entangled photons    play  a  crucial  role  in future  quantum  technologies, ranging from photonic quantum computing~\cite{RevModPhys.79.135,PhysRevLett.93.040503,PhysRevLett.95.010501,PhysRevLett.95.010501,Knill:2001aa} and communication~\cite{Azuma:2015aa,Li:2019aa,Buterakos2017,PhysRevX.10.021071,hilaire2020resource} to fundamental tests of quantum mechanics~\cite{Pan:2000aa,Lu.2014,PhysRevA.61.022109}. Several approaches for the generation of such multiphoton entangled states exist. One particular method relies on the well established technique for Bell state production via spontaneous parametric downconversion~(SPDC)~\cite{PhysRevLett.25.84,PhysRevLett.75.4337,PhysRevLett.83.3103}. The Bell pairs can subsequently be joined into larger photonic states using quantum states fusion~\cite{PhysRevLett.82.1345,PhysRevA.73.022330,PhysRevLett.78.3031}. This approach is, however, inherently probabilistic and thus limited to entangling only a modest number of photons~\cite{Lu:2007aa,Yao:2012aa,PhysRevLett.117.210502,PhysRevLett.121.250505}. 

A highly promising direction for deterministic generation of large entangled states is to exploit a single quantum emitter efficiently coupled to light to directly produce entangled photons in a sequential manner~\cite{PhysRevA.58.R2627,PhysRevA.61.062311,PhysRevLett.95.110503,PhysRevLett.103.113602,Lee_2019}. In the proof-of-principle experiment by Schwartz~\emph{et al.}~\cite{Schwartz434}, it was inferred that entanglement between five subsequent polarization-encoded photons could be  emitted by a single quantum dot. 
This founding experiment was conducted in a non-optimized setting using metastable dark excitons as qubits and without implementing photonic nanostructures. Thus, the entangled states produced so far do not have sufficient quality to allow for  all of the many envisioned applications.
It is thus essential to understand the mechanisms affecting the quality of the produced states in order to determine how well this system can be scaled up for generating multiple high-fidelity qubits.

\begin{figure}[t]	
\includegraphics[width=0.9\columnwidth]{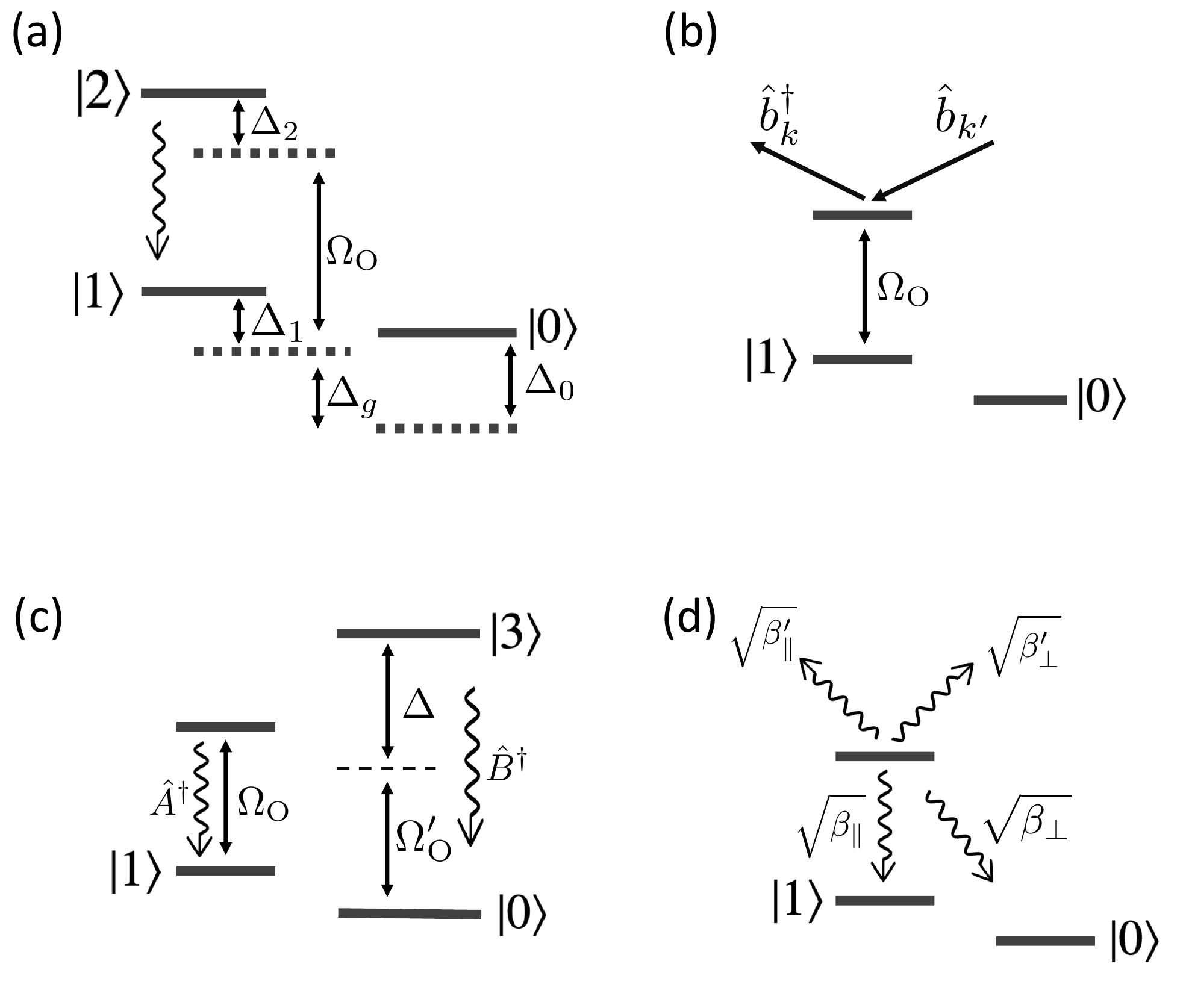}
\caption{\label{fig:2} Sources of imperfection. Errors considered include (a)~level shifting induced by interaction with the nuclear spin bath, (b)~phonon-induced pure dephasing, (c)~second order emissions from the resonant and far-detuned levels, and (d)~branching errors due to alternative decay paths. See text for explanations of each of the effects}. 
\end{figure}

In this paper, we perform a detailed theoretical analysis of a protocol for the generation of the time-bin entangled multiphoton states from a single quantum emitter~\cite{Lee_2019}, which in the  ideal situation can be described by the scheme shown in Fig.~\ref{fig:1}. Such an ideal scenario is, however, always corrupted by imperfections that inevitably occur in real physical systems. We consider multiple sources of errors, which are shown in Fig.~\ref{fig:2} and include (i)~ground-state dephasing, (ii)~phonon-induced pure dephasing, (iii)~excitation errors, and (iv)~photon emission errors. We derive simple analytical expressions for evaluating the fidelity of the produced entangled states for a given physical system. This theoretical understanding of the imperfections can then be used to optimize 
experimental realizations both in terms of efficiency and quality of the produced states.

For concreteness we consider two types of multi-photon entangled states, Greenberger–Horne–Zeilinger~(GHZ) and one-dimensional cluster states. The states will consists of $N$ photons entangled with a single spin. For convenience  we will label the state by the number of photons such that the $N$ photon GHZ state will have the form 
\begin{equation}
\label{eq:GHZ_model}
\ket{\mathrm{GHZ}^{(N)}} = \frac{1}{\sqrt{2}}
\Big{(}
\ket{0}^{\otimes N+1} + \ket{1}^{\otimes N+1}
\Big{)},
\end{equation}
which is a generalized version of Bell states to arbitrary number of particles. Here $\ket{0}$ and $\ket{1}$ denote logical states of the qubits. Being distributed over a network, this state allows for several interesting multi-user quantum protocols and thus serves a crucial resource for quantum network applications~\cite{PhysRevA.59.1829}.

Cluster states have attracted a lot of attention as a universal resource for one-way quantum computation~\cite{RevModPhys.79.135,PhysRevLett.93.040503,PhysRevLett.95.010501,PhysRevLett.95.010501,Knill:2001aa}, and lately also as a promising resource for  quantum repeaters~\cite{Azuma:2015aa,Li:2019aa,PhysRevX.10.021071}. In general cluster states can be obtained from arrays of qubits prepared in $|\Psi_0\rangle=(|0\rangle+|1\rangle)/\sqrt{2}$, by performing control phase gates between neighboring qubits along each of the dimensions of the cluster state. Unlike the GHZ state, cluster states do not allow for a compact-form expression since the number of terms grows rapidly with the number of qubits, but a two-qubit linear cluster state reads
\begin{equation}
    \begin{aligned}
    \label{eq:2_qubit_cluster}
    \ket{\mathrm{Cluster}^{(1)}} = 
    \frac{1}{2}
    \Big{(}
    &\ket{00}
    +
    \ket{01}
    +
    \ket{10}
    -
    \ket{11}
    \Big{)}.
    \end{aligned}
\end{equation}
This state can be transformed into $\ket{\mathrm{GHZ}^{(1)}}$ using local unitary operation, but for $N>1$ the state is in general much more complex.  
Most applications require two-dimensional cluster state, which in theory can be achieved by e.g. making use of coupled emitters~\cite{PhysRevLett.105.093601,PhysRevLett.123.070501} or fusing multiple linear cluster states~\cite{PhysRevA.91.042301}. 
As a starting point, we focus in this work on the generation of linear cluster states, which can be achieved with a single emitter  using the scheme in Fig.~\ref{fig:1}.

The reminder of this paper is organized as follows. In Sec.~\ref{sec:protocol}, we describe an idealized experimental protocol and introduce effective single-mode photon creation operators. In Sec.~\ref{sec:measurement}, we devise a theoretical formalism for calculation of the fidelities of the generated states. In Sec.~\ref{sec:fidelity}, we identify the main sources of imperfections expected to appear in real solid-state systems and derive expressions for corresponding infidelities. We assess the states fidelities for realistic experimental parameters and conclude with  future perspectives in Sec.~\ref{sec:conclusion}.

\section{Ideal protocol}\label{sec:protocol}
\begin{figure}[t]	
\includegraphics[width=0.85\columnwidth]{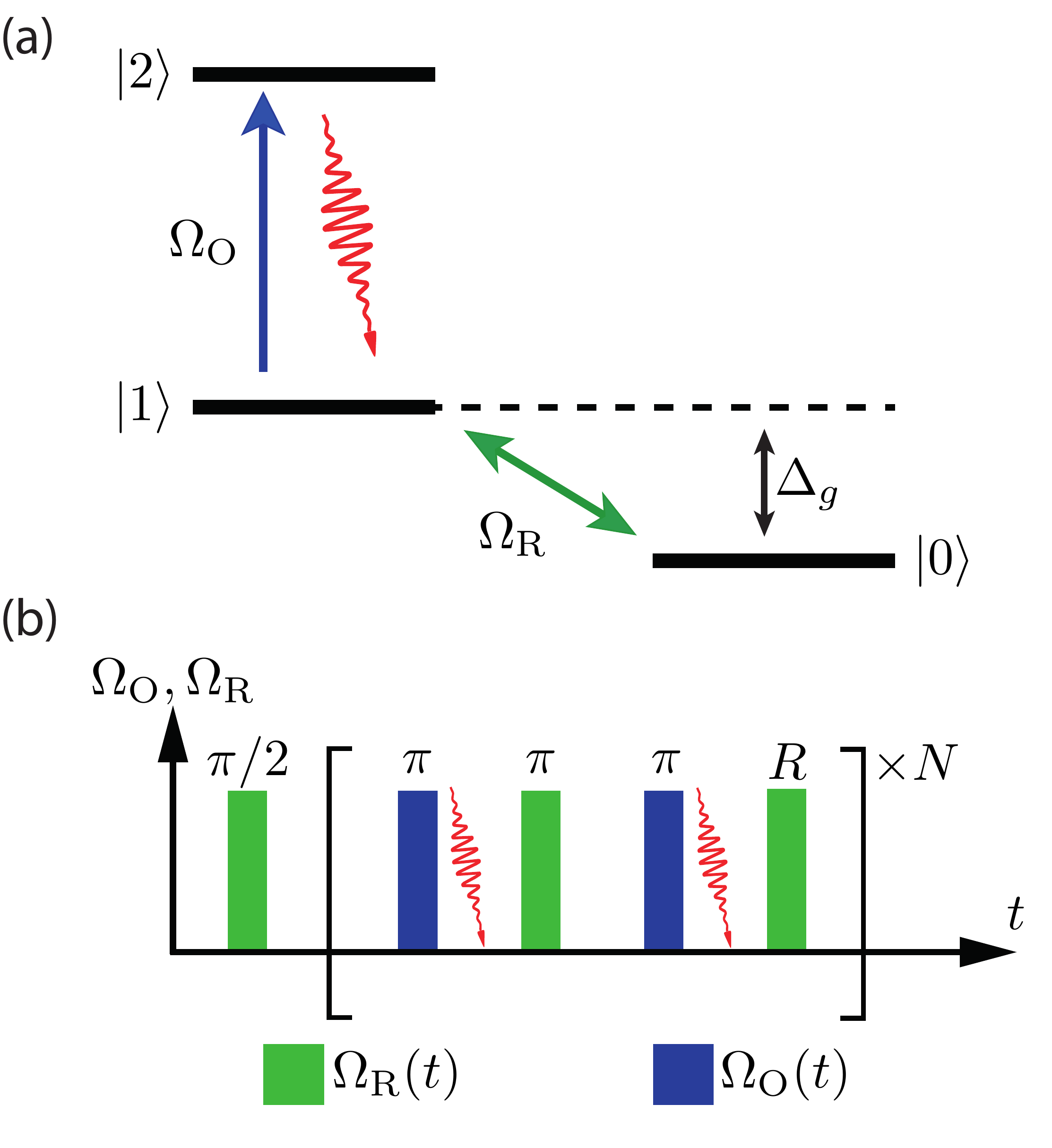}
\caption{\label{fig:1} Ideal protocol. (a)~Idealized level structure and (b)~driving pulse sequence  for the generation of the GHZ states~($\hat{R} = \hat{X}$) or cluster states~($\hat{R} = \hat{H}$). Laser pulses $\Omega_{\mathrm{O}}$ (green) and $\Omega_{\mathrm{R}}$ (blue) are used for driving the optical transition $\ket{1} \leftrightarrow \ket{2}$ and for the ground-state rotations $\ket{0} \leftrightarrow \ket{1}$, respectively. Following each optical $\pi$-pulse, we wait for the excited state $\ket{2}$ to decay back to the ground state $\ket{1}$ (red wiggly lines). Using $N$ repetitions of the pulse sequence, an entangled state of $N$ photons and the quantum emitter is generated.}
\end{figure}   
We begin with an idealized scheme proposed by Lee~\emph{et al.}~\cite{Lee_2019} that uses a periodically driven quantum emitter for the sequential generation of photons entangled in their relative arrival times. The use of the time-bin degrees of freedom to encode and transfer quantum information is highly attractive since it is ideally suited for distribution through optical fibers. Furthermore, the scheme offers a number of advantages for quantum dots, but we expect that it will also be very suitable for other quantum emitters such as atoms in cavities~\cite{Keller:2004aa,Barros_2009,PhysRevLett.120.203601} or color centers in diamond~\cite{Mizuochi:2012aa,Aharonovich_2011,Vasconcelos:2020aa}. The scheme for the sequential generation of time-bin entangled photons is illustrated in Fig.~\ref{fig:1} and goes as follows:
\begin{enumerate}
    \item The ground-state spin is initialized in the state $\ket{\Psi_0}=(\ket{0}+\ket{1})/\sqrt{2}$ using a $\pi/2$-pulse on the  $\ket{0} \leftrightarrow \ket{1}$ transition using the field $\Omega_{\mathrm{R}}$.
    \item The $\ket{1} \leftrightarrow \ket{2}$ transition is resonantly driven by a $\pi$-pulse with the field $\Omega_{\mathrm{O}}$, which generates a photon in an early time bin~$\ket{e}$ upon emission.
    \item The ground states $\ket{0}$ and $\ket{1}$ are flipped.
    \item Step 2 is repeated to generate a photon in a late time bin~$\ket{l}$.
    \item 
        \begin{itemize}
            \item GHZ state: step 3 is repeated.
            \item Cluster state: the Hadamard gate (or, equivalently, a $\pi/2$ rotation around the $x$ or $y$ axis) between the ground states $\ket{0}$ and $\ket{1}$ is applied.
        \end{itemize} 
    \item Steps 2--5 are repeated $N$ times to create an $N$-photon entangled state. 
\end{enumerate}

Following steps 2--5 for the GHZ state, the initial state transforms as
\begin{equation}
\begin{aligned}
\label{eq:ideal_protocol}
&\frac{1}{\sqrt{2}}
(\ket{0,\emptyset}
+
\ket{1,\emptyset})
\xrightarrow[]{2} 
\frac{1}{\sqrt{2}}
(\ket{0,\emptyset}
+
\ket{1,e})
\\&
\xrightarrow[]{3} 
\frac{1}{\sqrt{2}}
(\ket{1,\emptyset}
+
\ket{0,e})
\xrightarrow[]{4}
\frac{1}{\sqrt{2}}
(\ket{1,l}
+
\ket{0,e})
\\&
\xrightarrow[]{5} 
\frac{1}{\sqrt{2}}
(\ket{1,e}
+
\ket{0,l})
=
\frac{1}{\sqrt{2}}
(\hat{A}_{e}^{\dagger}\ket{1,\emptyset}
+
\hat{A}_{l}^{\dagger}\ket{0,\emptyset}),
\end{aligned}
\end{equation}
where $\ket{\emptyset}$ denotes the photon vacuum and the operator $\hat{A}_{e}^{\dagger}$~($\hat{A}_{l}^{\dagger}$) creates a single photon in an early~(late) time bin. Repeated $N$ times, the protocol produces an $N+1$ particle GHZ state of the form~\eqref{eq:GHZ_model} containing  $N$ photons and the spin. Here the spin state $\ket{0}$~($\ket{1}$) and the photon state $\ket{l}$~($\ket{e}$) are used as logical states $\ket{0}$~($\ket{1}$). Replacing the $\pi$-pulse in the Step 5 with the Hadamard gate produces the state
$(\ket{0,l} + \ket{0,e} + \ket{1,e} - \ket{1,l})/2$, which is identical to Eq.~\eqref{eq:2_qubit_cluster}. For higher $N$, the state is more complicated to write down, but we prove in Appendix~\ref{sec:Appendix_0} that the sequence produces a 1D-cluster state.

In the idealized protocol~\eqref{eq:ideal_protocol} described above, we do not go into details about the shape of the emitted photons. Taking a finite lifetime $1/\gamma$ of the excited state into account, the evolution during photon emission in the $(u,j)$th time interval can in a suitable rotating frame and under the Markov approximation be described by
\begin{equation}
    \begin{aligned}
    \label{eq:finite_emission}
    \ket{1,\emptyset}
    &\rightarrow
    \sqrt{\gamma}
    \int_0^{\infty}\textrm{d} t_e e^{-\frac{\gamma}{2}t_e}
    \hat{a}_{u,j}^{\dagger}(t_e)
    \ket{1,\emptyset},
    \end{aligned}
\end{equation}
while the state $\ket{0,\emptyset}$ stays intact. Each time bin is labelled by indices $(u,j)$, which correspond to the $j$th photon emitted in an early ($u=e$) or a late ($u=l$) part of the protocol. The operator $\hat{a}^{\dagger}_{u,j}(t_e)$ creates a photon at time $t_e$ during the $(u,j)$th time interval. In principle the integral in Eq.~\eqref{eq:finite_emission} should not go to infinity since we will have a finite duration $T/2$ of the early and late time bin. We assume, however, that $\gamma T \gg 1$ so that we can ignore exponentially small terms $\exp(-\gamma T/2)$ and extend the limit of the integration to infinity. 
Thus, the states after a single round of the protocol transform as in Eq.~\eqref{eq:ideal_protocol}, with photon creation operators $\hat{A}^{\dagger}_{u,j}$ taking the form
\begin{equation}
    \begin{aligned}
    \label{eq:single_mode_operator}
    \hat{A}^{\dagger}_{u,j,\mathrm{id}}
    =
    \sqrt{\gamma}
    \int_0^{\infty}\textrm{d} t_e e^{-\frac{\gamma}{2}t_e}\hat{a}_{u,j}^{\dagger}(t_e),
    \end{aligned}
\end{equation}
which obey the correct bosonic operators commutation relations, $[\hat{A}_{u,j},\hat{A}^{\dagger}_{u^{\prime},j^{\prime}}] = \delta_{u,u^{\prime}}\delta_{j,j^{\prime}}$.

For convenience, we define an ideal \emph{single-round} operator
\begin{equation}
    \begin{aligned}
    \label{eq:single-protocol_operators_ideal}
    \hat{O}^{\dagger}_{j,\mathrm{id}} &= 
    \hat{R}\Big{(}\ket{1}\bra{0} 
    \hat{A}_{l,j,\mathrm{id}}^{\dagger} 
    + 
    \ket{0}\bra{1} 
    \hat{A}^{\dagger}_{e,j,\mathrm{id}}\Big{)},
    \end{aligned}
\end{equation}
which corresponds to a single round of the protocol and, being applied to the spin state $\ket{\Psi_0}$, generates the $j$th photon in either the GHZ~($\hat{R}=\hat{X}$) or the cluster~($\hat{R}=\hat{H}$) state. The conventional notations $\hat{X}$ and $\hat{H}$ are here used to denote the Pauli-X and Hadamard gates. The ideal $N$-photon states therefore read
\begin{equation}
    \begin{aligned}
    \label{eq:ideal_state_general}
        \ket{\Psi_{\mathrm{id}}^{(N)}}
        =
        \hat{O}^{\dagger}_{N,\mathrm{id}}
        ..
        \hat{O}^{\dagger}_{1,\mathrm{id}}
        \ket{\Psi_0}\ket{\emptyset},
    \end{aligned}
\end{equation}
where $\ket{\Psi_0}$ is the initial spin state and $\ket{\emptyset} = \ket{\emptyset_1..\emptyset_N}$ is the $N$-photon vacuum. 

In a realistic situation the generation process will introduce errors and imperfections. We will take this into account by modifying the single-round operator~\eqref{eq:single-protocol_operators_ideal}, which in the most general case reads
\begin{equation}
    \begin{aligned}
    \label{eq:single-protocol_operators}
    \hat{O}^{\dagger}_{j} &= 
    \ket{1}\bra{0} \hat{A}^{\dagger}_{10,j} 
    +
    \ket{0}\bra{1} \hat{A}^{\dagger}_{01,j}
    \\&+
    \ket{0}\bra{0} \hat{A}^{\dagger}_{00,j} 
    +
    \ket{1}\bra{1} \hat{A}^{\dagger}_{11,j}.
    \end{aligned}
\end{equation}
{{Here $\hat{A}^{\dagger}_{kl,j}$ are general operators expressing the emission of photons for an emitter starting the $j$th period in state $\ket{l}$ and ending in state $\ket{k}$. The operators $\hat{A}^{\dagger}_{kl,j}$ contain all possible changes in the environment and the resulting leakage of information, e.g. due to phonon scattering or loss of photons during the pulse sequence. The  environmental  degrees  of  freedom  are  subsequently traced out when calculating the fidelity. This approach is slightly different than the typical master equation formalism, in which one  traces over the environment from the beginning to achieve a reduced density matrix for the system. The difference between the two approaches is, however, only at which stage one traces over the environmental degrees of freedom and their results are equivalent.
}}

In Sec.~\ref{sec:fidelity}, we take into account one imperfection at a time by constructing the corresponding operators~\eqref{eq:single-protocol_operators} and calculate its effect on the quality of the produced state.

\section{Entanglement characterization}\label{sec:measurement}

\subsection{Operational fidelity}\label{subsec:fidelity}
Before moving to the sources of imperfections, we briefly describe the experimental measurement process and introduce the corresponding measure of how ideal the state is. The measure of closeness between two states is conventionally given by the fidelity, which is defined as
\begin{equation}
\begin{aligned}
    \label{eq:strict_fidelity}
    \mathcal{F}^{(N)}_{\mathrm{exact}}
    &=
    \mathrm{Tr}_{\mathrm{env}}
    \Big{\{}
    \bra{\Psi_{\mathrm{id}}}
    \hat{\rho}^{(N)}
    \ket{\Psi_{\mathrm{id}}}
    \Big{\}}
    \\&=
    \mathrm{Tr}_{\mathrm{env}}
    \Big{\{}
    \bra{\Psi_{\mathrm{id}}}
    \hat{O}_N^{\dagger}
    ..
    \hat{O}_1^{\dagger}
    \ket{\Psi_0,\emptyset}
    \bra{\Psi_0,\emptyset}
    \hat{O}_1
    ..
    \hat{O}_N
    \ket{\Psi_{\mathrm{id}}}
    \Big{\}}
    ,
\end{aligned}
\end{equation}
where $\ket{\Psi_{\mathrm{id}}}$ is the ideal state~\eqref{eq:ideal_state_general} and $\hat{\rho}^{(N)}$ is the output $N$-photon state produced by the operators~\eqref{eq:single-protocol_operators}. The trace over environment here corresponds to any unobserved degree of freedom, e.g. emitted phonons or lost photons.

\begin{figure}[t]	
\includegraphics[width=0.9\columnwidth]{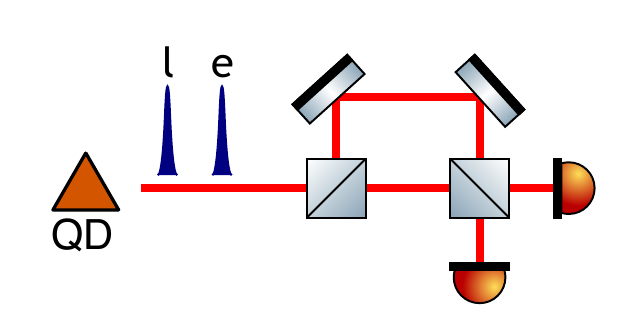}
\caption{\label{fig:3}Measurement setup for detecting time-bin entangled photons. Measurements are made in either the Z-basis by passing everything through the short arm, or in the X-basis by passing  early photons through the long arm and  late photons through the short arm of the  interferometer.} 
\end{figure}

Equation~\eqref{eq:strict_fidelity} compares the produced state with an outgoing photon in a well defined temporal mode. In most experimental situations, however, one does not have complete information about the temporal mode.  We will therefore slightly modify the strict definition of the fidelity~\eqref{eq:strict_fidelity} and introduce an \emph{operational fidelity}.
The typical experimental method of measuring time-bin encoded qubits is to interfere a photon pulse with a time delayed pulse as shown in Fig.~\ref{fig:3}. In the experiment, one distinguishes only between  early and a late time bins, while the exact time of photon emission within each time bin is not resolved or discarded in the analysis. Thus, one effectively has two sets of indices labeling time: the number of the time bin $j$ and the emission time within the time bin, $t_j$. Since the emission time is not used, we trace it out and obtain the \emph{operational fidelity},
\begin{widetext}
\begin{equation}
    \begin{aligned}
    \label{eq:exp_fidelity}
    \mathcal{F}^{(N)}_{}
    &=
    \mathrm{Tr}_{\mathrm{env}}
    \Big{\{}
    \int_0^{\infty}
    \mathrm{d} t_N
    ..
    \int_0^{\infty}
    \mathrm{d} t_1
    \bra{\emptyset}
    \bra{\Psi_0}
    \hat{o}_1(t_1)
    ..
    \hat{o}_N(t_N)
    \hat{\rho}^{(N)} 
    \hat{o}^{\dagger}_N(t_N)
    ..
    \hat{o}^{\dagger}_1(t_1)
    \ket{\Psi_0}
    \ket{\emptyset}
    \Big{\}}
    \\&=
    \mathrm{Tr}_{\mathrm{env}}
    \Big{\{}
    \int_0^{\infty}
    \mathrm{d} t_N
    ..
    \int_0^{\infty}
    \mathrm{d} t_1
    \bra{\emptyset}
    \bra{\Psi_0}
    \hat{o}_1(t_1)
    ..
    \hat{o}_N(t_N)
    \hat{O}^{\dagger}_N
    ..
    \hat{O}^{\dagger}_1
    \ket{\Psi_0}
    \ket{\emptyset}
    \bra{\emptyset}
    \bra{\Psi_0}
    \hat{O}_1
    ..
    \hat{O}_N
    \hat{o}^{\dagger}_N(t_N)
    ..
    \hat{o}^{\dagger}_1(t_1)
    \ket{\Psi_0}
    \ket{\emptyset}
    \Big{\}},
    \end{aligned}
    \end{equation}
\end{widetext}
where $\hat{\rho}^{(N)} = \ket{\Psi^{(N)}}\bra{\Psi^{(N)}}$ is the real state defined in~\eqref{eq:strict_fidelity} and the operators
\begin{equation}
    \begin{aligned}
    \label{eq:ideal_projectors}
    \hat{o}_{j}^{\dagger}(t_j) 
    &= 
    \hat{R}
    \Big{(}
    \ket{1}\bra{0}\hat{a}_{l,j}^{\dagger}(t_j) + \ket{0}\bra{1}\hat{a}_{e,j}^{\dagger}(t_j)
    \Big{)}
    \end{aligned}
\end{equation}
are the projectors on the ideal GHZ~($\hat{R} = \hat{X}$) or cluster~($\hat{R} = \hat{H}$) states. 

These two fidelity expressions~\eqref{eq:strict_fidelity} and~\eqref{eq:exp_fidelity} will in general give different results. Which of them provides a better description of concrete quantum information protocol will depend on the measurement performed in the specific protocol. If all photons are measured with a setup as in Fig.~\ref{fig:3}, then the fidelity in Eq.~\eqref{eq:exp_fidelity} provides a better description, whereas Eq.~\eqref{eq:strict_fidelity} may be a better choice if a different measurement sequence is used. 
As a specific example, the quantum repeater protocol of Ref.~\cite{PhysRevX.10.021071} considers photon numbers $N$ in the range 200--300. Out of these only a single photon is interfered with a different quantum emitter, whereas the remaining $N-1$ photons are measured in a setup as in Fig.~\ref{fig:3}. For this reason and since this is the experimentally most accessible quantity, we shall in the remainder of this article only consider the fidelity in Eq.~\eqref{eq:exp_fidelity}.

\subsection{Effects of photon loss and filtering}\label{subsec:filtering}

The definition of the operational fidelity~\eqref{eq:exp_fidelity} is yet to be modified in order to correspond to an experimentally realistic measurements. 

\emph{Photon losses} --- Successful detection of the emitted photons is limited by the collection of the photons from the waveguide, the subsequent propagation loss, and the detector efficiency. 
Due to these imperfections, experiments involving optical photons will have a nonzero probability to lose photons and only a fraction $\eta < 1$ of the produced photons will result in the detection event.  We model loss of a photon by modifying the single-mode creation operator~\eqref{eq:single_mode_operator} as
\begin{equation}
    \begin{aligned}
    \label{eq:freq_filter_single}
    \hat{A}^{\dagger} \rightarrow \sqrt{\eta}\hat{A}^{\dagger} + \sqrt{1-\eta}\hat{\tilde{A}}^{\dagger},
    \end{aligned}
\end{equation}
where $\hat{\tilde{A}}^{\dagger}$ corresponds to the photons that do not reach the detector. In an experimental realization optical loss would lead to cases of unsuccessful entanglement generation and detection. 

{{Photon loss is a major obstacle in most optical quantum information protocols. Therefore  realistic schemes for quantum information processing involving single photons are designed to have built-in correction procedures against photon loss, see e.g. Refs.~\cite{PhysRevLett.115.020502,PhysRevX.10.021071}. By post-selecting events where the correct number of photons are detected, unsuccessful photon detections are discarded in the quantum protocols and do not influence the fidelity of the successfully generated photons. We are therefore interested in computing the fidelity conditioned on the  detection of a photon in each cycle. This corresponds to projecting the output state on the detected photon subspace, i.e. $\hat{O}^{\dagger}_j \rightarrow \hat{P}_{n_j>0}\hat{O}^{\dagger}_j$ with $\hat{P}_{n_j>0} = \hat{1} - \ket{\emptyset}\bra{\emptyset}$. The probability of having a photon in each cycle of the protocol then reads
\begin{equation}\label{eq:P_success_general}
    P(n_1>0,...,n_N>0) 
    = 
    \mathrm{Tr}
    \Big{\{}
    \hat{P}_{n_1>0}..\hat{P}_{n_N>0}
    \hat{\rho}^{(N)}
    \Big{\}}.
\end{equation}
Since we only take into account experimental realizations with nonzero measurements in each cycle of the protocol, the probability to accept an experimental realization will decrease, hence decreasing the probability of a successful outcome. The conditional fidelity is then given by normalising $\mathcal{F}^{(N)}$ to the total success probability~\eqref{eq:P_success_general},
\begin{equation}
    \begin{aligned}
    \label{eq:conditional_fidelity}
    \tilde{\mathcal{F}}^{(N)}_{}
    =
    \frac{{\mathcal{F}}^{(N)}_{}}
    {\mathrm{Tr}
    \Big{\{}
    \hat{P}_{n_1>0}..\hat{P}_{n_N>0}
    \hat{\rho}^{(N)}
    \Big{\}}}.
    \end{aligned}
\end{equation}

The conditional fidelity above captures the quality of the state once the photons are successfully detected. The overall success probability does, however, influence the  performance of any quantum protocol. For instance  Ref. ~\cite{PhysRevX.10.021071} describe a quantum repeater protocol, which can in principle work for any single photon  efficiency above 50\%  but assumes 95\% for good performance, whereas the scheme for universal optical quantum computation in Ref. ~\cite{PhysRevLett.115.020502} tolerates a loss rate of 1.6\%. These numbers are challenging to achieve but solid state implementations are beginning to reach this level with a recent experiment reaching 85\% efficiency~\cite{PhysRevLett.123.183602,Bhaskar2020}. The efficiency is a separate issue from the quality and we here focus on the quality and evaluate the conditional fidelity. Alternatively the unconditional fidelity can be obtained from our results by simply multiplying the results with the associated success probability.}}

\emph{Temporal filtering} --- 
In the ideal protocol~\eqref{eq:ideal_protocol} described above, the excitation $\ket{1}\rightarrow\ket{2}$ was considered to be instantaneous. 
Realistically, transferring population to the excited state takes the  time of the Rabi $\pi$-cycle, which depends on the temporal shape and duration of the driving pulse. During this period photons from the driving laser can also leak into the detection arm of the setup. Thus, it is desirable to ignore the photons which were possibly emitted during the driving pulse or directly came from the laser. The undesired photons can be filtered out with near unit efficiency by keeping the detectors off during the driving laser pulse or by having shutters which only admits photons after the end of the excitation pulse.

\emph{Frequency filtering} --- The real systems have more complicated energy level structure than the scheme shown in Fig.~\ref{fig:1}(a). Not only the resonant transition is possible, but also emission of a photon from the far-detuned transitions, e.g. as in Fig.~\ref{fig:2}(c). In general, different transitions can have different collection efficiencies, which we take into account by redefining single-mode creation operators as
\begin{equation}
    \begin{aligned}
    \label{eq:freq_filter}
    \hat{A}^{\dagger} \rightarrow \sqrt{\eta_2}\hat{A}^{\dagger} + \sqrt{1-\eta_2}\hat{\tilde{A}}^{\dagger},
    \\
    \hat{B}^{\dagger} \rightarrow \sqrt{\eta_3}\hat{B}^{\dagger} + \sqrt{1-\eta_3}\hat{\tilde{B}}^{\dagger},
    \end{aligned}
\end{equation}
where $\eta_2 = \eta \xi_2$ and $\eta_3 = \eta \xi_3$ and the creation operators $\hat{A}^{\dagger}$ and $\hat{B}^{\dagger}$ correspond to the correct~(resonant) and the undesired~(off-resonant) photons, respectively. Without additional filtering, the photons from both transitions are collected with equal efficiency, i.e. $\xi_2 = \xi_3 = 1$. Cavity frequency filters can be added to the experimental setup in order to filter out the undesired photons and pass only the photons coming from the main transition. We will assume non-perfect filtering efficiency by applying filters such that $0 < \xi_3 \ll \xi_2 = 1$. Note that such (imperfect) frequency filtering is still compatible with perfect temporal filtering, e.g., if temporal shutters are placed before  a frequency filter.  

In the reminder of this paper, we take into account different imperfections present in real systems by modifying the corresponding single-protocol operators~\eqref{eq:single-protocol_operators} and calculating the fidelity according to Eqs.~(\ref{eq:exp_fidelity},\ref{eq:conditional_fidelity},\ref{eq:freq_filter}).

\section{Fidelity assessment}\label{sec:fidelity}

\subsection{Spin-state preserving errors}\label{subsec:spin_preserving}

We start our analysis by considering the errors that do not affect the spin states, but merely modify the single-mode creation operators~\eqref{eq:single_mode_operator} leaving the structure of the operators~\eqref{eq:single-protocol_operators_ideal} unaffected. These include ground- and excited-state dephasing, two-photon emission, and excitation of the off-resonant transitions.

Inserting the GHZ-state operators~\eqref{eq:single-protocol_operators} and~\eqref{eq:ideal_projectors} into~\eqref{eq:exp_fidelity} and assuming that the excitation at different time bins are uncorrelated, we obtain the unconditional fidelity of the GHZ state for non-spin-mixing errors,
\begin{equation}
\begin{aligned}
\label{eq:GHZ_non_mixing_fidelity}
    &\mathcal{F}^{(N)}_{}[\mathrm{GHZ}]
    \\&=
    \frac{1}{4}
    \mathrm{Tr}_{\mathrm{env}}
    \sum_{u,v=e,l}
    \Big{(}
    \int_0^{\infty}
    \mathrm{d} t
    \bra{\emptyset}
    \hat{a}_{u}(t)
    \hat{A}_{u}^{\dagger}
    \ket{\emptyset}
    \bra{\emptyset}
    \hat{A}_{v}
    \hat{a}_{v}^{\dagger}(t)
    \ket{\emptyset}
    \Big{)}^N.
    \end{aligned}
\end{equation}
Analogously, the cluster state fidelity reads
\begin{equation}
\begin{aligned}
\label{eq:Cluster_non_mixing_fidelity}
    &\mathcal{F}^{(N)}_{}[\mathrm{Cl}]
    \\&=
    \mathrm{Tr}_{\mathrm{env}}
    \Big{(}
    \frac{1}{4}
    \sum_{u,v=e,l}
    \int_0^{\infty}
    \mathrm{d} t
    \bra{\emptyset}
    \hat{a}_{u}(t)
    \hat{A}_{u}^{\dagger}
    \ket{\emptyset}
    \bra{\emptyset}
    \hat{A}_{v}
    \hat{a}_{v}^{\dagger}(t)
    \ket{\emptyset}
    \Big{)}^N.
    \end{aligned}
\end{equation}
The diagonal terms~($u=v$) in Eqs.~\eqref{eq:GHZ_non_mixing_fidelity}~and~\eqref{eq:Cluster_non_mixing_fidelity} correspond to the $z$-basis measurement, while the off-diagonal terms~($u \neq v$) correspond to the $x$-basis measurement, as explained in Fig.~\ref{fig:3}. The expressions above are derived under an assumption that the creation operators $\hat{A}^{\dagger}$ at different time intervals commute. This assumption is valid as long as one considers coupling to a Markovian environment or a non-Markovian classical noise, such as drift of the magnetic field or instability in the driving laser. The detailed derivations of the expressions~(\ref{eq:GHZ_non_mixing_fidelity},\ref{eq:Cluster_non_mixing_fidelity}) and a discussion of their applicability are given in  Appendix~\ref{sec:Appendix_A}.

\subsubsection{Ground-state dephasing}\label{subsubsec:ground_state_dephasing}
In solid state emitters, both electron and hole spin states suffer from interaction with the nuclear spin bath, an effect also referred to as the Overhauser noise~\cite{PhysRevB.93.205429}. It results in a short spin coherence times $T_2^*$, which becomes a limiting factor for a number of quantum information processing applications. Effectively, the spin-bath induced noise adds a random shift $\Delta_{i}$ to the energy levels, as shown in Fig.~\ref{fig:2}(a). The corresponding perturbation of the three-level Hamiltonian is given by
\begin{equation}
    \begin{aligned}
    \label{eq:three_level_H}
    \hat{H}^{\prime} &= 
    \sum_{i=0}^2 \Delta_i
    \ket{i}\bra{i}.
    \end{aligned}
\end{equation}
Writing a wavefunction ansatz as~\cite{das2019wavefunction} 
\begin{equation}
    \begin{aligned}
    \label{eq:WF_ansatz_3level}
    \ket{\Psi(t)}
    &= 
    c_2(t)
    \ket{2,\emptyset}
    +
    c_1(t)
    \ket{1,\emptyset}
    +
    c_0(t)
    \ket{0,\emptyset}
    \\&+
    \int_0^{\infty} \mathrm{d}t_e\phi(t,t_e)\hat{a}^{\dagger}(t_e)
    \ket{0,\emptyset}
    \end{aligned}
\end{equation}
and solving for coefficients $\phi_{2}(t,t_e), c_0(t)$ yields
\begin{equation}
    \begin{aligned}
    \label{eq:3level_coef}
    \phi(t,t_e) &= \sqrt{\gamma}e^{-i\Delta_1(t-t_e)
    }e^{-i\Delta_2 t_e}e^{-\frac{\gamma}{2}t_e}\theta(t-t_e)
    \\
    c_0(t) &= e^{-i\Delta_0 t}.
    \end{aligned}
\end{equation}
Thus, the states after the first half of the protocol transform according to
\begin{equation}
    \begin{aligned}
    \label{eq:finite_emission_T2*}
    \ket{1,\emptyset}
    &\rightarrow
    \sqrt{\gamma}
    e^{-i\frac{\Delta_1 T}{2}}
    \int_0^{\infty}\textrm{d} t_e e^{-\frac{\gamma}{2}t_e}
    e^{-i\Delta_{21} t_e}
    \hat{a}_{u,j}^{\dagger}(t_e)
    \ket{1,\emptyset},
    \\
    \ket{0,\emptyset}
    &\rightarrow
    e^{-i\frac{\Delta_0 T}{2}}
    \ket{0,\emptyset},
    \end{aligned}
\end{equation}
where we denote $\Delta_{21} = \Delta_2 - \Delta_1$. Note that after a single full round of the protocol, both $\ket{0}$ and $\ket{1}$ states will accumulate a global phase $e^{-i (\Delta_0 + \Delta_1)T/2}$, which is not important and can be omitted. The initial state $\ket{\Psi_0}$ therefore transforms according to the ideal protocol~\eqref{eq:ideal_protocol}, where the single-mode operators $\hat{A}^{\dagger}_{u,j}$ are defined as
\begin{equation}
\label{eq:single-mode_operator_Overhauser}
    \hat{A}^{\dagger}_{u,j}
    =
    \sqrt{\gamma}
    \int_0^{\infty}\textrm{d} t e^{-\frac{\gamma}{2}t}
    e^{- i\Delta_{21}t}
    \hat{a}_{u,j}^{\dagger}(t).
\end{equation}
With these operators, each of the four terms in Eqs.~\eqref{eq:GHZ_non_mixing_fidelity} and \eqref{eq:Cluster_non_mixing_fidelity} yields
\begin{equation}
    \begin{aligned}
    &\int_0^{\infty}
    \mathrm{d} t
    \bra{\emptyset}
    \hat{a}_{u}(t)
    \hat{A}_{u}^{\dagger}
    \ket{\emptyset}
    \bra{\emptyset}
    \hat{A}_{v}
    \hat{a}_{v}^{\dagger}(t)
    \ket{\emptyset}
    \\&=
    \gamma
    \int_0^{\infty}\mathrm{d} t
    \int_0^{\infty}\textrm{d} t^{\prime} 
    \int_0^{\infty}\textrm{d} t^{\prime\prime}
    \underbrace{\bra{\emptyset}
    \hat{a}_{u}(t)
    \hat{a}_{u}^{\dagger}(t^{\prime})
    \ket{\emptyset}}_{\delta(t - t^{\prime})}
    \\&
    \underbrace{
    \bra{\emptyset}
    \hat{a}_{v}(t^{\prime\prime})    
    \hat{a}_{v}^{\dagger}(t)
    \ket{\emptyset}}_{\delta(t - t^{\prime\prime})}
    e^{-\frac{\gamma}{2}(t^{\prime}+t^{\prime\prime})}
    e^{i\Delta_{12}(t^{\prime}-t^{\prime\prime})}=1,
    \end{aligned}
\end{equation}
and therefore
\begin{equation}
    \label{eq:Overhauser_fidelity}
    \boxed{
    \mathcal{F}^{(N)}_{T_2^*}[\mathrm{GHZ}] = \mathcal{F}^{(N)}_{T_2^*}[\mathrm{Cl}] = 1.}
\end{equation}

Strikingly, dephasing induced by the interaction with the nuclear spin bath or any other slow drift of the energy levels does not affect the quality of the produced state. This happens due to two reasons. First, as was pointed out earlier, the duration of the early and late parts of the protocol are equal, resulting in common global phase $e^{-i(\Delta_0 + \Delta_1)T/2}$, which we omit. This is reminiscent to a spin echo built into the time-bin generation protocol~\cite{PhysRev.80.580,PhysRevLett.100.236802,PhysRevLett.109.237601}. Second, the experiment does not resolve the exact photon emission time, but only the number of a time bin, i.e., the change from the fidelity definition in Eq.~\eqref{eq:strict_fidelity} to the fidelity in Eq.~\eqref{eq:exp_fidelity}. Here we only interfere photons which are emitted exactly $T/2$ apart using the interferometer in Fig.~\ref{fig:3}. This means  that the interfered photons come from events which have spent exactly the same time in the excited states, ensuring perfect spin echo conditions. The immediate consequence of Eq.~\eqref{eq:Overhauser_fidelity} in the context of quantum dot emitters is that while the coherence time of a hole spin is considerably longer than that of an electron, the dephasing is effectively removed in the protocol and both the electron spin and the hole spin can be used as the ground state qubit with equally good performance. 

Above we used the fact that the drift of the energy levels due to the Overhauser effect happens on the timescales much slower than a single round of the protocol and thus can be neglected. On longer times, however, such drift can potentially influence the coherence and is often referred to as $T_2$ noise. 
However, in our protocol the $\pi$-pulses are periodically applied in the middle of each repetition of the protocol, which has been shown to increase $T_2$ to few microseconds, thus suppressing the corresponding noise~\cite{PhysRevB.97.241413,Press2010} even if the number of produced photons is scaled to hundreds for typical quantum dot emission time scales. We thus expect slow drifts of energy levels to have a negligible effect on the quality of the produced multi-photon states and do not consider this type of dephasing.

\subsubsection{Phonon-induced pure dephasing}\label{subsubsec:phonon_induced_pure dephasing}
The next imperfection we study is pure dephasing of the excited state induced by scattering of phonons, as shown in Fig.~\ref{fig:2}(b). While the spin is excited, it can scatter phonons thereby inducing a random phase change at a rate $\gamma_{\mathrm{d}}$. The wavefunction corresponding to a single emitted phonon and one scattered phonon can be described as
\begin{equation}
    \begin{aligned}
    \label{eq:WF_with_photons}
    \ket{\Psi}_{u,j}
    &=
    \int_0^{\infty} 
    \mathrm{d} t_e
    \Big{(}
    \phi(t_e) 
    \\&+ 
    \sum_{k} \phi_k(t_e) \hat{b}^{\dagger}_{u,j,k}
    \Big{)}
    \hat{a}^{\dagger}_{u,j}(t_e)
    \ket{1,\emptyset,\tilde{\emptyset}},
    \end{aligned}
\end{equation}
where $\ket{\tilde{\emptyset}}$ denotes the vacuum state of phonons, $u = \{e,l\}$, and $j$ is the photon number. The operator $\hat{b}_{u,j,k}^{\dagger}$ creates a phonon in mode $k$ and $\bra{\tilde{\emptyset}} \hat{b}_{u,j,k} \hat{b}_{u^{\prime},j^{\prime},k^{\prime}}^{\dagger} \ket{\tilde{\emptyset}} = \delta_{u,u^{\prime}}\delta_{j,j^{\prime}}\delta_{k,k^{\prime}}$, i.e. we make a Markovian approximation for the phononic reservoir  such that phonons scattered in different time bins or different modes are orthogonal. In Eq.~\eqref{eq:WF_with_photons}, we only model a first-order scattering process and neglect the probabilities to scatter more than one phonon per cycle. Since the scattering of even a single phonon will  remove all coherence with the excited states, the scattering of a second phonon will not further reduce the fidelity and it is sufficient to consider the scattering of a single. 
The coefficients in Eq.~\eqref{eq:WF_with_photons} were derived in Ref.~\cite{Mikkel.Msc.2019} and read
\begin{equation}
    \begin{aligned}
    \label{eq:phonon_coefficients}
    \phi(t_e) & = \sqrt{\gamma} 
    e^{- (\frac{\gamma}{2} + \gamma_{\mathrm{d}})t_e} 
    \\
    \sum_k
    |\phi_{k}(t_e)|^2 
    &=
    \gamma
    e^{-\gamma t_e}
    \Big{(}
    1 - e^{-2\gamma_{\mathrm{d}}t_e}
    \Big{)}.
    \end{aligned}
\end{equation}

\begin{figure}[t]
\includegraphics[width=0.9\columnwidth]{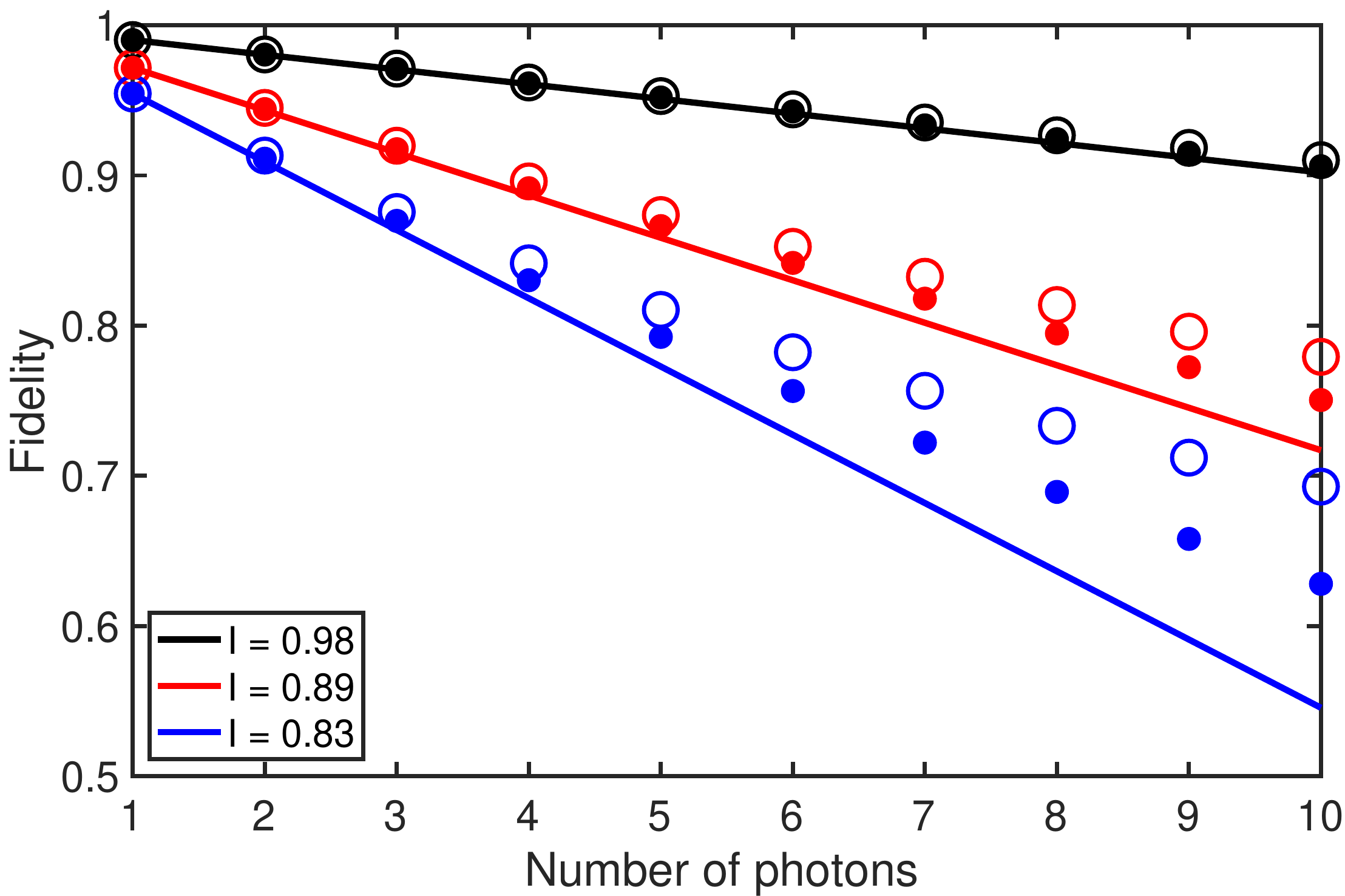}
\caption{\label{fig:4} Fidelities of the GHZ~(circles) and the cluster~(dots) states in the presence of phonon dephasing. The first-order approximation~\eqref{eq:fidelity_with_phonons_approx} for both the GHZ and the cluster state is shown with solid lines and agrees well with the exact solution~\eqref{eq:fidelity_with_phonons} for a large fidelity $F\gtrsim 0.8$. Black, red, and blue curves correspond to the dephasing rates $\gamma_{\mathrm{d}}/\gamma = 0.01$,
$\gamma_{\mathrm{d}}/\gamma = 0.03$, and 
$\gamma_{\mathrm{d}}/\gamma = 0.05$, respectively.} 
\end{figure}

As in the case of ground-state dephasing,  phonon scattering does not alter the spin states and merely modifies the single-mode operators~\eqref{eq:single_mode_operator} which now read
\begin{equation}
\begin{aligned}
    \label{eq:single-mode_operator_phonons}
    \hat{A}^{\dagger}_{u,j} 
    = 
    \int_0^{\infty} \textrm{d} t_e 
    \Big{(} 
    \phi(t_e)
    +
    \sum_{k} \phi_{k}(t_e)
    e^{-i\omega_k t_e}
    \hat{b}^{\dagger}_{u,j,k} 
    \Big{)}
    \hat{a}^{\dagger}_{u,j}(t_e).
\end{aligned}
\end{equation}

Substituting single-mode operators~\eqref{eq:single-mode_operator_phonons} and the coefficients~\eqref{eq:phonon_coefficients} into the off-diagonal terms of  Eqs.~\eqref{eq:GHZ_non_mixing_fidelity} and \eqref{eq:Cluster_non_mixing_fidelity} yields
\begin{equation}
    \begin{aligned}
    \label{eq:phonon_offdiagonal_term}
        &\mathrm{Tr}_{\mathrm{ph}}
        \Big{\{}
        \int_0^{\infty}
        \mathrm{d} t
        \bra{\emptyset}
        \hat{a}_{u}(t)
        \hat{A}_{u}^{\dagger}
        \ket{\emptyset,\tilde{\emptyset}}
        \bra{\emptyset,\tilde{\emptyset}}
        \hat{A}_{v}
        \hat{a}_{v}^{\dagger}(t)
        \ket{\emptyset}
        \Big{\}}
        \\&=
        \int_0^{\infty}
        \mathrm{d} t
        |\phi(t)|^2
        =
        \frac{\gamma}{\gamma+2\gamma_{\mathrm{d}}},
    \end{aligned}
\end{equation}
where $u \neq v$. In Eq.~\eqref{eq:phonon_offdiagonal_term} only the terms that do not contain the phonon creation operators $\hat{b}^{\dagger}$ survive since phonons scattered in an early and a late time bins are orthogonal. The diagonal terms of Eqs.~\eqref{eq:GHZ_non_mixing_fidelity} and \eqref{eq:Cluster_non_mixing_fidelity} with the coefficients~\eqref{eq:phonon_coefficients} become unity,
\begin{equation}
    \begin{aligned}
    \label{eq:phonon_diagonal_term}
        &\mathrm{Tr}_{\mathrm{ph}}
        \Big{\{}
        \int_0^{\infty}
        \mathrm{d} t
        \bra{\emptyset}
        \hat{a}_{u}(t)
        \hat{A}_{u}^{\dagger}
        \ket{\emptyset,\tilde{\emptyset}}
        \bra{\emptyset,\tilde{\emptyset}}
        \hat{A}_{u}
        \hat{a}_{u}^{\dagger}(t)
        \ket{\emptyset}
        \Big{\}}
        \\&=
        \int_0^{\infty}
        \mathrm{d} t
        \Big{(}
        |\phi(t)|^2
        +
        \sum_k
        |\phi_{k}(t)|^2
        \Big{)}
        =
        1,
    \end{aligned}
\end{equation}
where no cross terms of the form $\phi(t)\phi_k^*(t)$ are present since $\mathrm{Tr}_{\mathrm{ph}}{\{} \hat{b}^{\dagger}_{u,j,k} \ket{\tilde{\emptyset}}\bra{\tilde{\emptyset}} {\}}=0$.

Finally, inserting~\eqref{eq:phonon_offdiagonal_term} and \eqref{eq:phonon_diagonal_term} into Eqs.~\eqref{eq:GHZ_non_mixing_fidelity} and \eqref{eq:Cluster_non_mixing_fidelity} yields the fidelity of the $N$-photon GHZ and cluster states in the presence phonon-induced pure dephasing,
\begin{equation}
    \label{eq:fidelity_with_phonons}
    \boxed{
    \begin{aligned}
    \mathcal{F}_{\mathrm{ph}}^{(N)}[\mathrm{GHZ}]
    &=
    \frac{1}{2} 
    + 
    \frac{1}{2}
    \Big{(}
    \frac{\gamma}{\gamma+2\gamma_{\mathrm{d}}}
    \Big{)}^N
    =
    \frac{1 + I^N}{2},
    \\
    \mathcal{F}_{\mathrm{ph}}^{(N)}[\mathrm{Cl}]
    &=
    \frac{1}{2^N} 
    \Big{(}
    1 + \frac{\gamma}{\gamma+2\gamma_{\mathrm{d}}}
    \Big{)}^N
    =
    \Big{(}\frac{1 + I}{2}\Big{)}^N,
    \end{aligned}}
\end{equation}
where the degree of indistinguishability is defined as $I = \gamma/(\gamma + 2\gamma_{\mathrm{d}})$~\cite{RevModPhys.87.347}. Since typically $\gamma_{\mathrm{d}} \ll \gamma$, the expressions above can be expanded around $\gamma_{\mathrm{d}}/(\gamma+2\gamma_{\mathrm{d}}) = 0$. In the first-order approximation, the fidelities of the two states become identical,
\begin{equation}
    \label{eq:fidelity_with_phonons_approx}
    \begin{aligned}
    {\mathcal{F}}_{\mathrm{ph,approx}}^{(N)}
    =
    1 - N\frac{\gamma_{\mathrm{d}}}{\gamma+2\gamma_{\mathrm{d}}}
    =
    1 - N\frac{1-I}{2}.
    \end{aligned}
\end{equation}
 Figure~\ref{fig:4} shows plots of the fidelities for a realistic range of parameters and different number of photons.

\subsubsection{Excitation errors}\label{subsubsec:excitation_errors}

Next, we take into consideration the errors that can occur during the excitation of the transition $\ket{1} \leftrightarrow \ket{2}$. The possible errors consist of two components. First is the probability of emitting a photon already during the finite duration of the driving laser pulse used to excite the $\ket{1} \leftrightarrow \ket{2}$ transition. In the discussion above, the excitation process was considered to be instantaneous and photons were only retrieved during the relaxation time of the protocol following the pump pulse. However, photon emissions during the driving pulse are possible and should be taken into account. We assume temporal filtering by keeping detectors inactive while driving the system to the excited state, as discussed in Sec.~\ref{subsec:filtering}. Hence photons emitted during the driving pulses are assumed to be lost and we only consider photons emitted during the subsequent period of free decay. 

A second source of imperfection considered here is the probability of exciting a far-detuned transition $\ket{0} \leftrightarrow \ket{3}$ as shown in Fig.~\ref{fig:2}(c). Cavity filters are assumed to suppress contributions from  off-resonant photons emitted on this transition, but the excitation of this will still induce dephasing due to multi-photon emission and the filtering may not be perfect. Hence we need to evaluate the effect of this.  

We note that the two effects depend on the temporal shape and length of the driving laser pulse in opposite ways: short high-intensity pulses would allow to highly suppress the second-order photon emission at the cost of strongly driving the undesired $\ket{0} \leftrightarrow \ket{3}$ transition. On the other hand, long and weak driving pulses can suppress the excitation of the off-resonant transition but will result in photon emission during the pulse. Thus, our goal here is two-fold: (i) to find an optimal regime of the driving laser and (ii) to evaluate the corresponding fidelity of multiphoton states. 

An extensive analysis of this system is provided in Ref.~\cite{Pol.Msc.2019}. 
Below we merely outline the central results. 
We start by writing a wavefunction ansatz as~\cite{das2019wavefunction} 
\begin{widetext}
    \begin{equation}
    \begin{aligned}
    \label{eq:WF_ansatz}
            \ket{\Psi_{eg}}
            &=
            c_g(t)\ket{g,\emptyset}
            +
            c_e(t)\ket{e,\emptyset}
            +
            \int \mathrm{d} t_e
            \phi_g(t,t_e)
            \hat{a}^{\dagger}(v_g(t-t_e))
            \ket{g,\emptyset}
            +
            \int \mathrm{d} t_e
            \phi_e(t,t_e)
            \hat{a}^{\dagger}(v_g(t-t_e))
            \ket{e,\emptyset}
            \\&+
            \int \mathrm{d} t_{e_1}
            \int \mathrm{d} t_{e_2}
            \phi_{gg}(t,t_{e_1},t_{e_2})
            \hat{a}^{\dagger}(v_g(t-t_{e_1}))
            \hat{a}^{\dagger}(v_g(t-t_{e_2}))
            \ket{g,\emptyset},
    \end{aligned}
    \end{equation}
\end{widetext}
where $\hat{a}^{\dagger}(z) = \frac{1}{\sqrt{2\pi}}\int \mathrm{d}k \hat{a}^{\dagger}_k e^{{i(k-k_0)z}}$ is the creation operator in real space and $v_g$ is the group velocity. The first three terms in Eq.~\eqref{eq:WF_ansatz} are analogous to the scenario described by Eq.~\eqref{eq:WF_ansatz_3level} and include excited and ground states amplitudes $c_g(t)$ and $c_e(t)$ and the first-order photon emission process~[$\phi_g$]. The wavefunction~\eqref{eq:WF_ansatz} furthermore considers the possibility to emit a photon during the pulse and to be re-excited~[$\phi_e$], and the possibility to emit one photon during and one photon after the pulse~[$\phi_{gg}$]. Since we currently do not consider the possibility of transitions between the two branches $\ket{1}\leftrightarrow\ket{2}$ and  $\ket{0}\leftrightarrow\ket{3}$, the wavefunction ansatz~\eqref{eq:WF_ansatz} can be written and solved separately for the resonant two-level system~($\{g,e\}=\{1,2\}$) and the undesired far-detuned transition~($\{g,e\}=\{0,3\}$). 

Taking into account all possible outcomes, the states upon photon emission become
\begin{equation}
        \begin{aligned}
        \label{eq:two_photon_state}
        \ket{1,\emptyset}
        &\rightarrow
        \Big{(}c_1 + c_2\hat{A}_0^{\dagger} + \Phi_1\hat{A}_{p_1}^{\dagger}
        +
        \Phi_2\hat{A}_{p_2}^{\dagger}\hat{A}_{0}^{\dagger}
        \Big{)}\ket{1,\emptyset},
        \\
        \ket{0,\emptyset}
        &\rightarrow
        \Big{(}
        c_0 + c_3\hat{B}_0^{\dagger} + \Phi_0\hat{B}_{p_1}^{\dagger}
        +
        \Phi_3\hat{B}_{p_2}^{\dagger}\hat{B}_{0}^{\dagger}
        \Big{)}
        \ket{0,\emptyset},
        \end{aligned}
\end{equation}
where the coefficients $\Phi_i$ are such that $
    |\Phi_i(T_{\mathrm{p}})|^2 = 
    v_g\int 
    \mathrm{d} t_e
    |\phi_i(T_{\mathrm{p}},t_e)|^2$.
The creation operators $\hat{A}^{\dagger}$~($\hat{B}^{\dagger}$) correspond to emission of a photon from the resonant $\ket{1} \leftrightarrow \ket{2}$~(off-resonant $\ket{0} \leftrightarrow \ket{3}$) transition during~($\hat{A}^{\dagger}_{p_i},\hat{B}^{\dagger}_{p_i}$) or after~($\hat{A}^{\dagger}_{0},\hat{B}^{\dagger}_{0}$) the pulse. For simplicity we have here ignored the possibility of two photons being emitted during the pulse. We therefore evaluate the wavefunction in Eq.~\eqref{eq:WF_ansatz} at the end of the pulse at time $T_{\mathrm{p}}$ keeping at most a single emission during the pulse. After  this time the system will emit a photon if it is in the excited states. This emission process is independent of the dynamics during the excitation process and is the denoted by the same operators $A_0$ and $B_0$ irrespective of the dynamics during the pulse. Furthermore all coefficients are to be evaluated at the end of the pulse $T_{\mathrm{p}}$. A single round of the protocol therefore corresponds to the action of an operator 
\begin{equation}
    \begin{aligned}
    \label{eq:single-protocol_operators_second_order}
    \hat{O}^{\dagger}_{j} &= \hat{R}\Big{(}\ket{1}\bra{0} \hat{Q}^{\dagger}_{l,j} + \ket{0}\bra{1} \hat{Q}^{\dagger}_{e,j}\Big{)},
    \end{aligned}
\end{equation}
which has the same form as~\eqref{eq:single-protocol_operators} with the single-mode operators $\hat{A}_{u,j}^{\dagger}$ replaced by the effective creation operators
\begin{equation}
    \begin{aligned}
    \label{eq:Q_operator}
    \hat{Q}^{\dagger}_{u,j} &= 
    \Big{(}
    c_1 + c_2\hat{A}_0^{\dagger,u} + \Phi_1\hat{A}_{p_1}^{\dagger,u}
    +
    \Phi_2\hat{A}_{p_2}^{\dagger,u}\hat{A}_{0}^{\dagger,u}
    \Big{)}_j
    \\&\times
    \Big{(}
    c_0 + c_3\hat{B}_0^{\dagger,v} + \Phi_0\hat{B}_{p_1}^{\dagger,v}
    +
    \Phi_3\hat{B}_{p_2}^{\dagger,v}\hat{B}_{0}^{\dagger,v}
    \Big{)}_j,
    \end{aligned}
\end{equation}
where $j$ is a photon number and $\{u,v\} = \{e,l\}$, $u \neq v$. 

The operator~\eqref{eq:Q_operator} includes all possible combinations of photons emitted from two two-level systems in a single round of the protocol. An ideal noiseless protocol corresponds to a single photon coming from the resonant transition, i.e. to $\hat{Q}_{u,\mathrm{ideal}}^{\dagger} = \hat{c}_2\hat{c}_0\hat{A}_{0}^{\dagger,u}$ with $|c_0| = |c_2| = 1$. 
In order to improve the quality of the produced state, we consider a combination of temporal and spectral filters to the output state~\eqref{eq:single-protocol_operators_second_order}, which suppress contributions from the terms other than $\hat{c}_2\hat{c}_0\hat{A}_0^{\dagger,u}$ in~\eqref{eq:Q_operator} as discussed in section~\ref{subsec:filtering}. 
First, we accept only  experimental instances which contain at least one photon emitted after the driving pulse, i.e. we reject  states that do not contain $\hat{A}^{\dagger}_0$ or $\hat{B}^{\dagger}_0$ in~\eqref{eq:single-protocol_operators_second_order} at each round of the protocol. Next, we use frequency filters to reject the photons emitted at the correct time, but with off-resonant frequency, that is, we suppress the contribution from the $\hat{B}_0^{\dagger}$ as described in Eq.~\eqref{eq:freq_filter}. 

The full state after applying the temporal and frequency filters to the state is given in Appendix~\ref{sec:Appendix_B}. The single-protocol operators~\eqref{eq:Q_operator} do not mix the spin states, and therefore the expressions for the fidelities~\eqref{eq:GHZ_non_mixing_fidelity} and \eqref{eq:Cluster_non_mixing_fidelity} are still valid, with photons emitted during the pulse playing the role of an environment. Since photons from different time bins are orthogonal, the only term that survives the trace operation in the off-diagonal terms of Eqs.~\eqref{eq:GHZ_non_mixing_fidelity} and~\eqref{eq:Cluster_non_mixing_fidelity} is
\begin{equation}
    \begin{aligned}
    \label{eq:offdiagonal_term_second_order}
        D_1 &= \mathrm{Tr}_{\mathrm{p}}
        \Big{\{}
        \int_0^{\infty}
        \mathrm{d} t
        \bra{\emptyset}
        \hat{a}_{v}(t)
        \hat{Q}^{\dagger}_{v,j}
        \ket{\emptyset}
        \bra{\emptyset}
        \hat{Q}_{u,j}
        \hat{a}_{u}^{\dagger}(t)
        \ket{\emptyset}
        \Big{\}}
        \\&=
        \eta_2|c_0c_2|^2
    \end{aligned}
\end{equation}
with $u \neq v$. The diagonal terms read
\begin{equation}
    \begin{aligned}
    \label{eq:diagonal_term_second_order}
        D_2 &= \mathrm{Tr}_{\mathrm{p}}
        \Big{\{}
        \int_0^{\infty}
        \mathrm{d} t
        \bra{\emptyset}
        \hat{a}_{u}(t)
        \hat{Q}^{\dagger}_{u,j}
        \ket{\emptyset}
        \bra{\emptyset}
        \hat{Q}_{u,j}
        \hat{a}_{u}^{\dagger}(t)
        \ket{\emptyset}
        \Big{\}}
        \\&=
        \eta_2
        \Big{(}
        |c_0c_2|^2
        +
        |c_0\Phi_2|^2
        +
        |\Phi_0c_2|^2
        +
        |\Phi_0\Phi_2|^2
        \Big{)}
        \\&+
        \eta_2(1-\eta_3)
        \Big{(}
        |c_2c_3|^2
        +
        |\Phi_2c_3|^2
        +
        |\Phi_3c_2|^2
        +
        |\Phi_3\Phi_2|^2
        \Big{)}.
    \end{aligned}
\end{equation}

Finally,  postselection is taken into account by dividing the fidelity by the success probability, i.e. the probability that at least one  photon has been detected,
\begin{equation}
    \begin{aligned}
    \label{eq:success_prob_second_order}
    P(n_j>0) = D_2 + D_3,
    \end{aligned}
\end{equation}
where 
\begin{equation}
\begin{aligned}
\label{eq:D3}
D_3
&=
\eta_3
\Big{(}
|c_3c_1|^2
+
|c_3\Phi_1|^2
+
|\Phi_3c_1|^2
+
|\Phi_3\Phi_1|^2
\\+&
|c_3c_2|^2
+
|c_3\Phi_2|^2
+
|\Phi_3c_2|^2
+
|\Phi_3\Phi_2|^2
\Big{)}.
\end{aligned}
\end{equation}

\begin{figure}[t]	
\includegraphics[width=0.9\columnwidth]{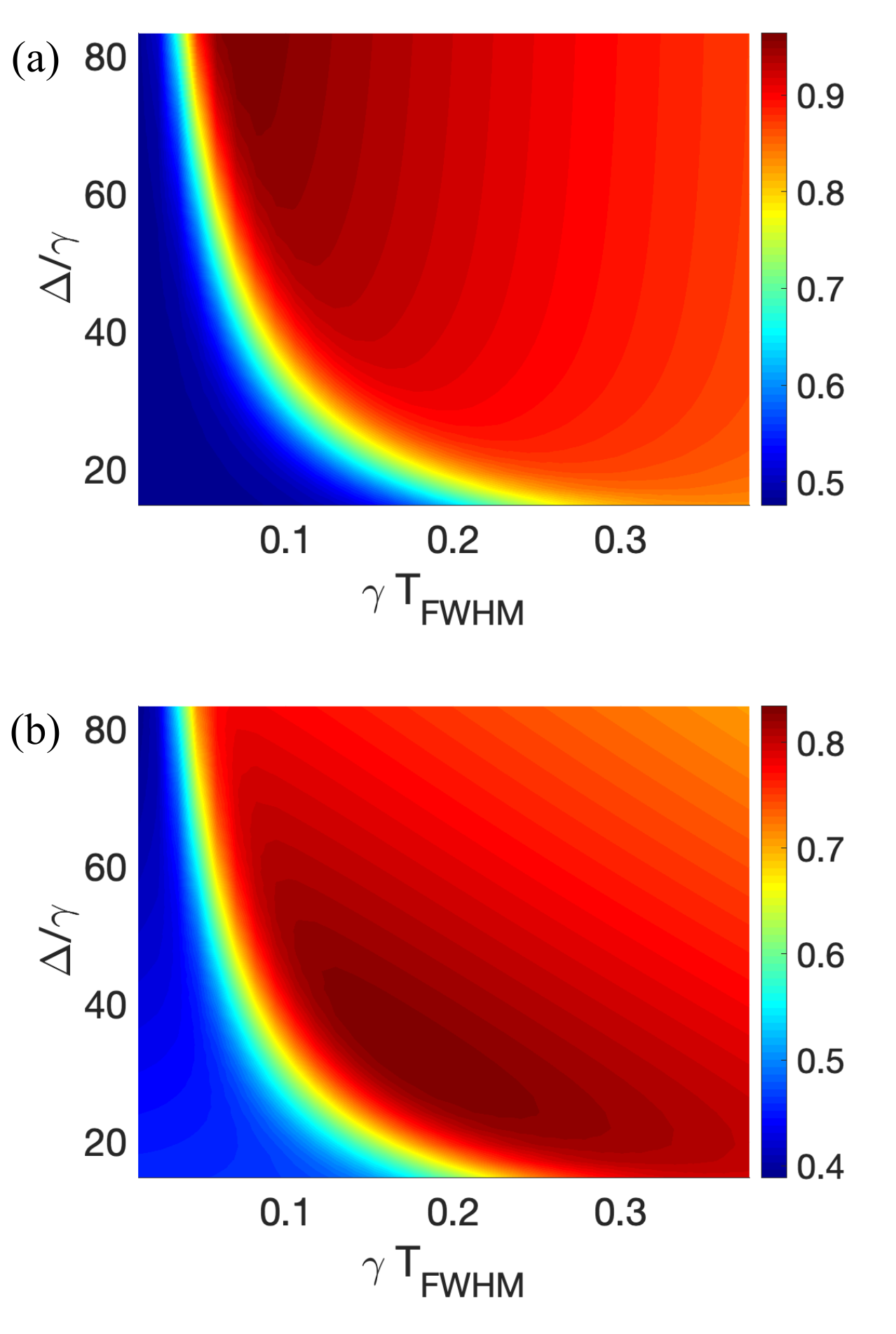}
\caption{\label{fig:5} Conditional fidelities of the five-photon GHZ state for a Gaussian-shaped pulse as a function of the pulse length $\gamma T_{\mathrm{FWHM}}$~($x$ axis) and  lifetime of the excited state, $\Delta/\gamma$~($y$ axis) at a fixed $\Delta = 2\pi \times 16$~$\mathrm{ns}^{-1}$. Panels (a) and (b) correspond to the fidelity with excitation errors $\tilde{\mathcal{F}}_{\mathrm{exc}}$ only and to the combined fidelity $\tilde{\mathcal{F}}_{\mathrm{comb}}$, respectively. In~(b) we have assumed a ratio of $\Delta/\gamma_{\mathrm{d}}=1.7 \times 10^3$, which roughly corresponds to quantum dots in a $2$~T magnetic field at a temperature of 1.8~K.}
\end{figure}

Substituting Eqs.~(\ref{eq:offdiagonal_term_second_order},\ref{eq:diagonal_term_second_order},\ref{eq:success_prob_second_order},\ref{eq:D3}) into~\eqref{eq:conditional_fidelity} yields the conditional fidelities of the GHZ and the cluster states,
\begin{equation}
    \label{eq:fidelity_with_second_order_emission}
    \boxed{
    \begin{aligned}
    \tilde{\mathcal{F}}_{\mathrm{exc}}^{(N)}[\mathrm{GHZ}]
    &=
    \frac{1}{2} 
    \frac{D_1^N + D_2^N}
    {(D_2+D_3)^N},
    \\
    \tilde{\mathcal{F}}_{\mathrm{exc}}^{(N)}[\mathrm{Cl}]
    &=
    \frac{1}{2^N}
    \Big{(}
    \frac{D_1 + D_2}
    {D_2+D_3}
    \Big{)}^N.
    \end{aligned}}
\end{equation}
The detailed derivations of Eqs.~(\ref{eq:offdiagonal_term_second_order},\ref{eq:diagonal_term_second_order},\ref{eq:success_prob_second_order},\ref{eq:D3}) are provided in~Appendix~\ref{sec:Appendix_B}.

The final step is the calculation of the coefficients in Eq.~\eqref{eq:fidelity_with_second_order_emission}, which depend on the temporal shape of the driving light pulse, photon emission rate $\gamma$, detuning $\Delta$, and filtering efficiencies $\xi_2, \xi_3$. In the Appendix~\ref{sec:Appendix_B}, we provide a system of coupled differential equations for the coefficients $\{c_i(t),\Phi_i(t)\}$, which where solved numerically under the assumption that the driving laser pulse has a Gaussian temporal profile. Figure~\ref{fig:5}(a) shows the calculated conditional fidelity of the five-photon GHZ state. We vary the emission rate $\gamma$ and full width half maximum pulse length $T_{\mathrm{FWHM}}$ while keeping a fixed detuning of the state $\ket{3}$ from the resonant transition. To ensure that we have a finite pulse duration, experiments will have to truncate the Gaussian pulse. In our simulation we do this at $T_\mathrm{p} = 3.2\times T_{\mathrm{FWHM}}$. Choosing a too long pulse duration will affect the success probability since the excitation in the excited state will decay, and hence a compromise will have to be made between the truncation of the Gaussian and the success probability.

\begin{figure}[t]	
\includegraphics[width=0.9\columnwidth]{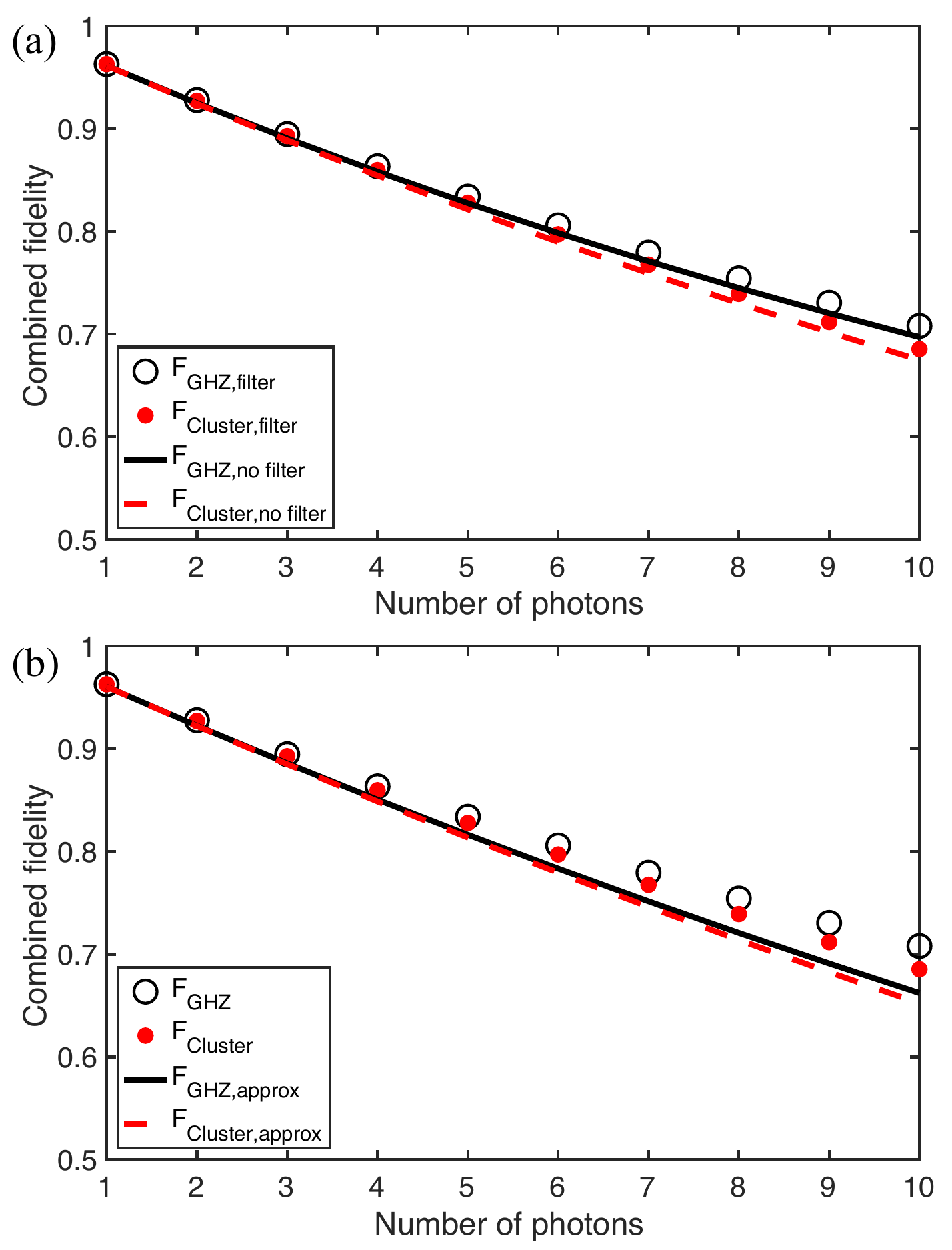}
\caption{\label{fig:6} 
(a)~Effect of frequency filtering.  Optimized combined fidelities of the GHZ~(circles) and the cluster~(dots) states as a function of a number of photons with frequency filters such that $\eta_2 = 1$, $\eta_3 = 0.02$. Solid and dashed lines correspond to, respectively, the GHZ and the cluster state without frequency filtering.
(b)~Validity of the first-order and square-shaped pulse approximations. Comparison between the combined fidelity of the GHZ state from the exact numerical solution for a Gaussian driving pulse~\eqref{eq:fidelity_with_second_order_emission}~(circles)
and its first-order approximation for a square pulse~\eqref{eq:excitation_with_filters_1st_order_text}~(black solid line), respectively. The exact and first-order fidelities of the cluster state are shown with red dots and dashed lines, respectively. Parameters values are given in the text.}
 \end{figure}

As follows from Fig~\ref{fig:5}(a), reducing the emission rate $\gamma$ at fixed values of the detuning $\Delta$ and $\gamma T_{\mathrm{FWHM}}$ improves the fidelity of the state. On the other hand, according to Eq.~\eqref{eq:fidelity_with_phonons}, increasing the emission rate results in a higher dephasing fidelity since the system spends less time in the excited state. Therefore, we show the combined fidelity $\tilde{\mathcal{F}}_{\mathrm{comb}} =  \mathcal{F}_{\mathrm{ph}}\tilde{\mathcal{F}}_{\mathrm{exc}}$ in Fig.~\ref{fig:5}(b) assuming fixed values $\Delta = 2\pi\times 16$~$\mathrm{ns}^{-1}$ and $\gamma_{\mathrm{d}} = 0.06$~$\mathrm{ns}^{-1}$. The calculated optimal parameters are $T_{\mathrm{FWHM,opt}} = 0.06$~ns and $\gamma_{\mathrm{opt}} = 3.2$~$\mathrm{ns}^{-1}$, which corresponds to a degree of indistinguishability $I_{\mathrm{opt}} = 0.96$ and falls within the experimentally realistic range of parameters for quantum dots. Figure~\ref{fig:5} only shows the results for the GHZ state since the fidelity of the cluster state is almost identical in the considered range of parameters. 
Figure~\ref{fig:6}(a) shows comparison between the fidelities with and without frequency filters. Evidently, frequency filters have a very small effect on the excitation errors in the assumed range of experimental parameters, but as shown below it will have a much larger effect on the branching error.

In  Appendix~\ref{sec:Appendix_B} we also derive an analytical solution corresponding to the simplified model where (i)~a square-shaped driving pulse is used, (ii)~frequency filters have perfect efficiency, $\xi_2 = 1$, $\xi_3 = 0$, and (iii) dynamics is solved up to the first order in perturbation theory. Furthermore we chose the pulse duration and intensity such that a perfect $\pi$-pulse is achieved on the resonant transition, whereas the off-resonant transition perform an off-resonant $2\pi$ Rabi oscillation, ideally returning all amplitude from state $\ket{3}$ to state $\ket{0}$. With these assumptions, the conditional fidelity reads
\begin{equation}
\begin{aligned}
\label{eq:excitation_with_filters_1st_order_text}
    \tilde{\mathcal{F}}^{(N)}_{\mathrm{exc,sq}} = 1 - N\frac{\sqrt{3}\pi}{8}\frac{\gamma}{\Delta}.
\end{aligned}
\end{equation}
Using a square shape approximation of the driving laser pulse serves as a good approximation and gives a simple analytical expression for the fidelity at optimal parameters, as one can observe in Fig.~\ref{fig:6}(b). 

\subsection{Branching error}\label{subsec:branching}

In the preceding discussion, we have assumed that only the vertical transitions Fig.~\ref{fig:2}(c) were allowed. This section is devoted to studying a more complex decay structure.

First, we study how imperfect decay ratios affect the quality of the generated quantum state. A level diagram with an additional decay path is shown in Fig.~\ref{fig:2}(d). Since the excitation of the level $|3\rangle$ in Fig.~\ref{fig:2}(c) constitute an error in itself, we will ignore a similar additional decay path from this level and focus on the level scheme in Fig.~\ref{fig:2}(d). We characterize the two decay paths by parameters $\beta_{\parallel}$ and $\beta_{\perp}$, which are the probabilities to emit a photon into the waveguide through the correct~($\ket{2} \xrightarrow{} \ket{1}$) or the incorrect diagonal~($\ket{2} \xrightarrow{} \ket{0}$) transition, respectively. 

Next, we consider photon losses, which we will divide into intrinsic and extrinsic losses. Intrinsic losses correspond to the photon emitted through the vertical or the diagonal transition out of the waveguide mode, corresponding to two additional processes shown in Fig.~\ref{fig:2}(d). The two processes occur with probabilities $\beta_{\parallel}^{\prime}$ and $\beta_{\perp}^{\prime}$ for the desired and undesired transition respectively. Extrinsic losses were discussed in Sec.~\ref{subsec:filtering} and correspond to the overall efficiency $\eta$ of the experimental setup and include all possible losses between the waveguide  and the detector. Taken together, the probabilities to emit and detect a photon coming from either the vertical or the diagonal transition are given by
\begin{equation}
    \begin{aligned}
    \label{eq:branching_coefficients_with_filters}
    p_{\parallel} &= \eta_2 \beta_{\parallel}
    \\
    p_{\perp} &= \eta_3 \beta_{\perp},
    \end{aligned}
\end{equation}
while the probabilities of losing the corresponding photons are
\begin{equation}
    \begin{aligned}
    \label{eq:branching_coefficients_with_filters_2}
    p_{\parallel}^{\prime} 
    &= 
    \beta_{\parallel}^{\prime} + (1-\eta_2) \beta_{\parallel}
    \\
    p_{\perp}^{\prime} 
    &=
    \beta_{\perp}^{\prime} + (1-\eta_3) \beta_{\perp},
\end{aligned}
\end{equation}
where $\eta_i = \eta \xi_i$, $\eta \le 1$, and $\xi_i=1$ ($\xi_i \ll 1$) without (with)  frequency filters on the $i$th transition. The full state after a single round of the protocol then reads
\begin{widetext}
    \begin{equation}
    \begin{aligned}
    \label{eq:single_round_branching}
    \ket{\Psi^{(1)}} 
    &= 
    \frac{1}{\sqrt{2}}
    \hat{R}
    \Big{(}
    \ket{0}
    \Big{\{}
    \sqrt{p_{\parallel}} 
    \ket{e, \emptyset} 
    +
    \sqrt{p^{\prime}_{\parallel}} 
    \ket{1_{e\parallel}, \emptyset}
    +
    \sqrt{p_{\perp}} 
    \ket{l^{\prime}, \emptyset} 
    +
    p_{\perp}
    \ket{l^{\prime}, e^{\prime}} 
    +
    \sqrt{p_{\perp} p^{\prime}_{\perp}}
    \ket{l^{\prime}, 1_{e\perp}} 
    +
    \sqrt{p_{\perp}^{\prime}}
    \ket{1_{l\perp}, \emptyset} 
    +
    \sqrt{p_{\perp}p^{\prime}_{\perp}}
    \ket{e^{\prime}, 1_{l\perp}} 
    \\&+
    p^{\prime}_{\perp}
    \ket{1_{e\perp}, 1_{l\perp}} 
    \Big{\}} 
    +
    \ket{1}
    \Big{\{}
    \sqrt{p_{\parallel}} 
    \ket{l, \emptyset} 
    +
    \sqrt{p_{\parallel} p_{\perp}} 
    \ket{l, e^{\prime}}
    +
    \sqrt{p_{\parallel} p^{\prime}_{\perp}} 
    \ket{l, 1_{e\perp}} 
    +
    \sqrt{p^{\prime}_{\parallel}}
    \ket{1_{l\parallel}, \emptyset} 
    +
    \sqrt{p^{\prime}_{\parallel} p_{\perp}} 
    \ket{e^{\prime},1_{l\parallel}}
    \\&+
    \sqrt{p^{\prime}_{\parallel}, p^{\prime}_{\perp}}
    \ket{1_{l\parallel}, 1_{e\perp}} 
    \Big{\}}
    \Big{)},
    \end{aligned}
    \end{equation}
\end{widetext}
where $\ket{e^{\prime}}$ and $\ket{l^{\prime}}$ are, respectively, early and late photons emitted into the waveguide through the diagonal $\ket{2} \rightarrow \ket{0}$ transition of Fig.~\ref{fig:2}(d) and $\ket{1_{u\parallel}}$~($\ket{1_{u\perp}}$) denotes a late~($u=l$) or an early~($u=e$) photon that has been lost after being emitted in a vertical~(diagonal) transition. Again the operator $\hat{R}$ is $\hat{X}$ and $\hat{H}$ for the GHZ and the cluster state, respectively. As expected, Eq.~\eqref{eq:single_round_branching} reduces to the ideal state~\eqref{eq:ideal_state_general} for $p_{\parallel} = 1$ and $p_{\perp} = p_{\perp}^{\prime} = p_{\parallel}^{\prime} = 0$.

The expressions for the fidelities~(\ref{eq:GHZ_non_mixing_fidelity},\ref{eq:Cluster_non_mixing_fidelity}) were derived under the assumptions that only the vertical transitions between spin states were allowed, and are thus not valid when imperfect branching in Fig.~\ref{fig:2}(d) is taken into account. We thus need to derive new expressions for the fidelity in this case. This calculation is different for the GHZ and cluster states, and will thus be handled separately below. 

\subsubsection{GHZ state with branching errors}\label{subsubsec:branching}

We start by calculating the fidelity of the GHZ state, which corresponds to $\hat{R}=\hat{X}$ in Eqs.~\eqref{eq:single_round_branching} and~\eqref{eq:ideal_projectors}. Using the same formalism as in the previous sections, the single round operators that produce a correct GHZ state read
\begin{equation}
    \begin{aligned}
    \label{eq:GHZ_single_round_branching}
    \hat{O}_j^{\dagger}
    &=
    \ket{1}\bra{1}
    \Big{(}
    \sqrt{p_{\parallel}}
    \hat{A}^{\dagger}_{e,j}
    +
    \sqrt{p_{\perp}^{\prime}p_{\perp}}
    \hat{A}^{\dagger}_{e^{\prime},j}
    \hat{L}_j^{\dagger}
    \Big{)}
    \\&+
    \ket{0}\bra{0}
    \sqrt{p_{\parallel}}
    \hat{A}^{\dagger}_{l,j}
    +
    \ket{0}\bra{1}
    \sqrt{p_{\perp}^{\prime}p_{\parallel}}
    \hat{A}^{\dagger}_{l,j}
    \hat{E}_j^{\dagger},
    \end{aligned}
\end{equation}

\noindent where $\hat{L}_j^{\dagger} = \ket{1_{l\perp,j}}\bra{\emptyset}$ and $\hat{E}_j^{\dagger} = \ket{1_{e\perp,j}} \bra{\emptyset}$. Therefore, for a single round of the protocol, 
\begin{equation}
    \begin{aligned}
    \label{eq:GHZ_single_photon_operator_br}
    \hat{o}_1(t_1)\hat{O}_1^{\dagger}
    &=
    \ket{1}\bra{1}
    \hat{a}_{e,1}(t_{1})
    \Big{(}
    \sqrt{p_{\parallel}}
    \hat{A}_{e,1}^{\dagger}
    +
    \sqrt{p_{\perp}^{\prime}p_{\perp}}
    \hat{A}_{e^{\prime},1}^{\dagger}
    \hat{L_1^{\dagger}}
    \Big{)}
    \\&+
    \ket{0}\bra{0}
    \sqrt{p_{\parallel}}
    \hat{a}_{1}(t_{1})
    \hat{A}_{l,1}^{\dagger}
    \\&+
    \ket{0}\bra{1}
    \sqrt{p_{\perp}^{\prime}p_{\parallel}}
    \hat{a}_{l,1}(t_{1})
    \hat{A}_{l,1}^{\dagger}
    \hat{E_1^{\dagger}}.
    \end{aligned}
\end{equation}
Repeating the protocol $N$ times with the initial spin state $\ket{\Psi_0}$, we arrive at 
    \begin{equation}
    \begin{aligned}
    \label{eq:GHZ_branching_N_operator}
    &\bra{\Psi_0}\hat{o}_1(t_1)..\hat{o}_N(t_N)\hat{O}_N^{\dagger}..\hat{O}_1^{\dagger}\ket{\Psi_0}
    \\&=
    \frac{1}{2}
    \Big{[}
    \prod_{j=1}^{N}
    \hat{a}_{e,j}(t_{j})
    \Big{(}
    \sqrt{p_{\parallel}}
    \hat{A}_{e,j}^{\dagger}
    +
    \sqrt{p_{\perp}^{\prime}p_{\perp}}
    \hat{A}_{e^{\prime}j}^{\dagger}
    \hat{L_j^{\dagger}}
    \Big{)}
    \\&+
    {p_{\parallel}}^{N/2}
    \Big{(}
    \prod_{j=1}^{N}
    \hat{a}_{l,j}(t_{j})
    \hat{A}_{l,j}^{\dagger}
    +
    \sqrt{p_{\perp}^{\prime}}
    \prod_{j=1}^{N}
    \hat{a}_{l,j}(t_{j})
    \hat{A}_{l,j}^{\dagger}
    \hat{E_j^{\dagger}}
    \Big{)}
    \Big{]}.
    \end{aligned}
\end{equation}

We now insert Eq.~\eqref{eq:GHZ_branching_N_operator} and its Hermitian conjugate into~\eqref{eq:exp_fidelity}~(for detailed derivations, see  Appendix~\ref{sec:Appendix_C}) and arrive at the expression for the unconditional fidelity of the GHZ state,
\begin{equation}
    \begin{aligned}
    \label{eq:fidelity_of_GHZ_branching_unconditional}
    \mathcal{F}^{(N)}_{\mathrm{br}}[\mathrm{GHZ}]
    =
    \frac{\Big{(}p_{\parallel} +
    {p_{\perp}^{\prime} p_{\perp}}\Big{)}^N
    +
    p_{\parallel}^N
    \Big{(}
    3
    +
    p_{\perp}^{\prime}\Big{)}}
    {4}.
    \end{aligned}
\end{equation}
Since we reject the experimental outcomes where no photons have been detected, the final conditional fidelity~\eqref{eq:conditional_fidelity} has to be normalized to the probability of detection  $P(n_1>0,..n_{N}>0)$ and becomes
\begin{equation}
    \begin{aligned}
    \label{eq:fidelity_of_GHZ_branching_conditional}
    \tilde{\mathcal{F}}^{(N)}_{\mathrm{br}}[\mathrm{GHZ}]
    =
    \frac{\Big{(}p_{\parallel} +
    {p_{\perp}^{\prime} p_{\perp}}\Big{)}^N
    +
    p_{\parallel}^N
    \Big{(}
    3
    +
    p_{\perp}^{\prime}\Big{)}}
    {4P(n_1>0,..n_{N}>0)}.
    \end{aligned}
\end{equation}

Each round of the protocol mixes the spin states due to the branching error, and the probability of detecting a photon in each of $N$ rounds is not merely a product of individual probabilities. Instead, the success probability can be expanded as a product of conditional probabilities. 
Let $P(n_j,s=s_0)$ be a probability that a photon has been emitted and detected in $j$th round of the protocol with the spin ending in a state $s_0$. Then the following set of equations can be written:
\begin{widetext}
\begin{equation}
    \begin{aligned}
    \label{eq:branching_success_probability_N}
    &P(n_1,..n_{N}) 
    = 
    P(n_1,..n_{N},s=0)
    +
    P(n_1,..n_{N},s=1)
    =
    \begin{pmatrix} 1 & 1 \end{pmatrix}
    \begin{pmatrix}
    P(n_1,..n_{N},s=0)\\
    P(n_1,..n_{N},s=1)
    \end{pmatrix}
    \\&=
    \begin{pmatrix} 1 & 1 \end{pmatrix}
    \begin{pmatrix}
    P(n_{N},s=0|n_1,..n_{N-1},s=0)
    P(n_1,..n_{N-1},s=0)
    +
    P(n_{N},s=0|n_1,..n_{N-1},s=1)
    P(n_1,..n_{N-1},s=1)
    \\
    P(n_{N},s=1|n_1,..n_{N-1},s=0)
    P(n_1,..n_{N-1},s=0)
    +
    P(n_{N},s=1|n_1,..n_{N-1},s=1)
    P(n_1,..n_{N-1},s=1)
    \end{pmatrix}
    \\&=
    \begin{pmatrix} 1 & 1 \end{pmatrix}
    \begin{pmatrix}
    P(n_{1},s=0|n_1,..n_{N-1},s=0)
    &
    P(n_{N},s=0|n_1,..n_{N-1},s=1)
    \\
    P(n_{N},s=1|n_1,..n_{N-1},s=0)
    &
    P(n_{N},s=1|n_1,..n_{N-1},s=1)
    \end{pmatrix}
    \begin{pmatrix}
    P(n_1,..n_{N-1},s=0)
    \\
    P(n_1,..n_{N-1},s=1)
    \end{pmatrix}
    \\&=
    \begin{pmatrix} 1 & 1 \end{pmatrix}
    \begin{pmatrix}
    M_{00}
    &
    M_{01}
    \\
    M_{10}
    &
    M_{11}
    \end{pmatrix}
    \begin{pmatrix}
    P(n_1,..n_{N-1},s=0)
    \\
    P(n_1,..n_{N-1},s=1)
    \end{pmatrix}
    =
    \begin{pmatrix} 1 & 1 \end{pmatrix}
    M^N
    \begin{pmatrix}
    P(s=0)
    \\
    P(s=1)
    \end{pmatrix}
    =
    \begin{pmatrix} 1 & 1 \end{pmatrix}
    M^N
    \begin{pmatrix}
    1/2
    \\
    1/2
    \end{pmatrix},
\end{aligned}
\end{equation}
\end{widetext}
where the matrix $M$ consists of the elements $M_{ij}$ which are the probabilities to detect a photon while changing spin state from $j$ to $i$ between adjacent repetitions of the protocol. The elements of $M$ can be derived by taking into account all possible processes in~\eqref{eq:single_round_branching} that result in at least one photon detection,
\begin{equation}
\begin{aligned}
    \label{eq:M_elements_GHZ}
    M_{00} &= p_{\parallel} \\
    M_{11} &= p_{\parallel} + 2p_{\perp}p^{\prime}_{\perp} + p^{2}_{\perp} \\
    M_{01} &= p_{\parallel}(p_{\perp} + p^{\prime}_{\perp}) + p_{\parallel}^{\prime}p_{\perp}\\
    M_{10} &= p_{\perp}.
\end{aligned}
\end{equation}

\begin{figure}[t]	
\includegraphics[width=0.9\columnwidth]{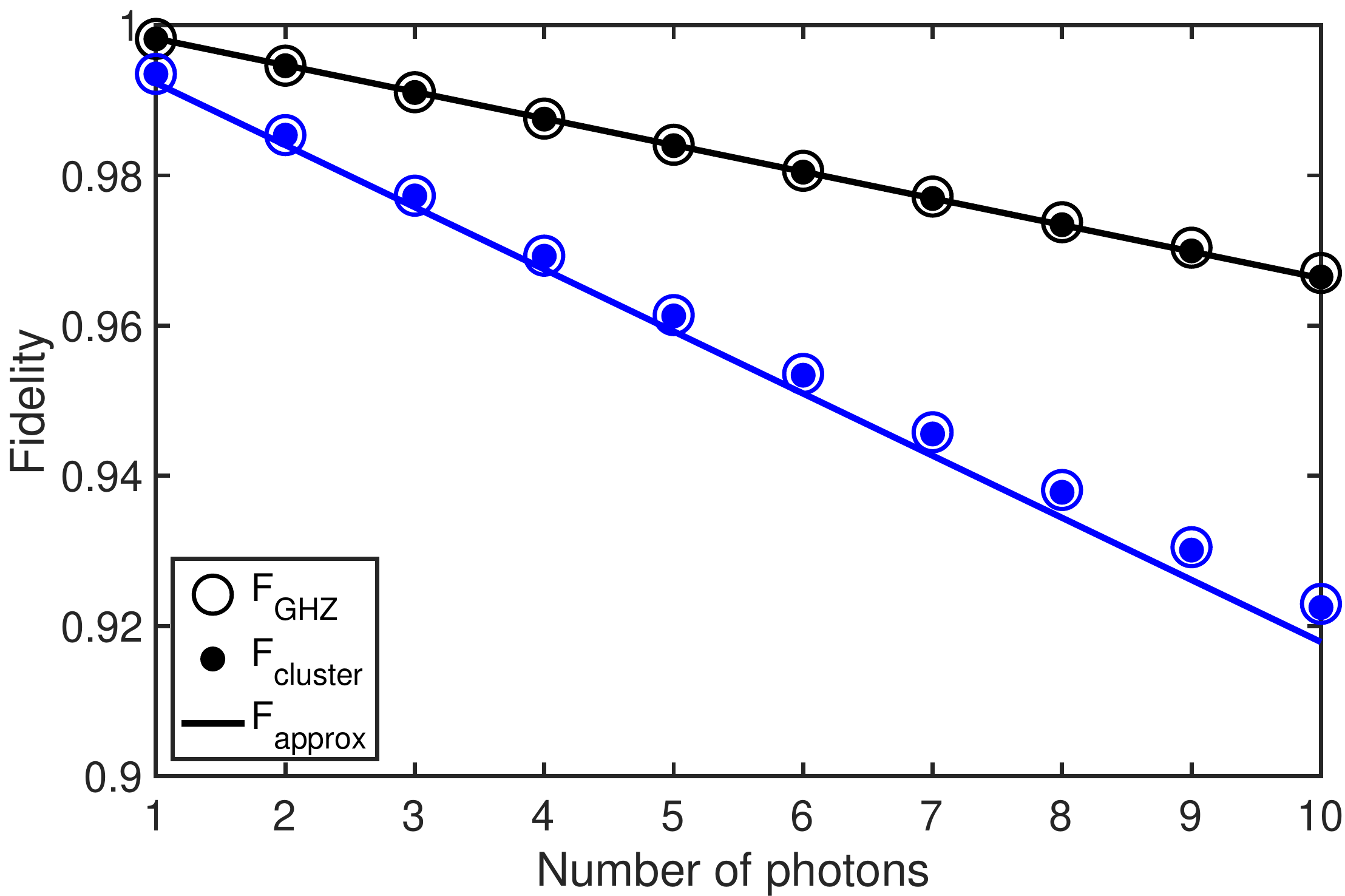}
\caption{\label{fig:7} State fidelities with imperfect branching. Numerically calculated fidelities of the GHZ and cluster states are shown circles and dots, respectively. The solid lines show the first order perturbative expressions for both the GHZ and the cluster states. Black and blue colors correspond to the fidelities with and without frequency filters, respectively. Here we use numerically simulated branching parameters for quantum dots embedded in a photonic crystal waveguide~\cite{Tiurev2020b}: $\beta_{\perp} = 0.05$, $\beta_{\perp}^{\prime} = 0.0025$, $\beta_{\parallel}^{\prime} = 0.0025$, $\beta_{\parallel} = 1 - \beta_{\perp} - \beta_{\perp}^{\prime} - \beta_{\parallel}^{\prime} = 0.99$, corresponding to a branching ratio of $B \approx 140$.}
\end{figure}

Finally, we insert~(\ref{eq:fidelity_of_GHZ_branching_unconditional},\ref{eq:branching_success_probability_N},\ref{eq:M_elements_GHZ}) into Eq.~\eqref{eq:conditional_fidelity} to obtain an expression for the conditional fidelity. Numerically calculated conditional fidelities for different number of photons are shown in Fig.~\ref{fig:7}. 

The success probability~\eqref{eq:branching_success_probability_N} can be expanded up to the first order around a small parameter $p_{\mathrm{wrong}}/p_{\parallel} \ll 1$ as
\begin{equation}\nonumber
\begin{aligned}
&P^{(N)}
=
\frac{1}{2}
\begin{pmatrix}
1 & 1
\end{pmatrix}
M^N
\begin{pmatrix}
1 \\ 1
\end{pmatrix}
\\&=
\frac{p_{\parallel}^N}{2}
\begin{pmatrix}
1 & 1
\end{pmatrix}
\Big{[}
\hat{1}
+
\frac{1}{p_{\parallel}}
\begin{pmatrix}
0 & p_{\parallel}(p_{\perp} + p_{\perp}^{\prime})+ p_{\perp}p_{\parallel}^{\prime} \\
p_{\perp} & p_{\perp}(p_{\perp} + 2p_{\perp}^{\prime})
\end{pmatrix}
\Big{]}^N
\begin{pmatrix}
1 \\ 1
\end{pmatrix}
\\&=
p_{\parallel}^N
\Big{(}
1
+
\frac{N}{2p_{\parallel}}
\Big{[}
p_{\parallel}(p_{\perp} + p_{\perp}^{\prime})
+
p_{\parallel}^{\prime}p_{\perp}
+
p_{\perp}(1 + p_{\perp} + 2p_{\perp}^{\prime})
\Big{]}
\Big{)},
\end{aligned}
\end{equation}
where $p_{\mathrm{wrong}}$ is any $p$ other than $p_{\parallel}$. The conditional fidelity~\eqref{eq:conditional_fidelity}  to first order in $p_{\mathrm{wrong}}/p_{\parallel}$ then becomes
\begin{equation}
\begin{aligned}
\label{eq:GHZ_fidelity_branching_conditional}
    &\tilde{\mathcal{F}}^{(N)}[\mathrm{GHZ}]
    \\&=
    \frac{\mathcal{F}^{(N)}[\mathrm{GHZ}]}{p_{\parallel}^N
    \Big{(}
    1
    +
    \frac{N}{2p_{\parallel}}
    \Big{[}
    p_{\parallel}(p_{\perp} + p_{\perp}^{\prime})
    +
    p_{\parallel}^{\prime}p_{\perp}
    +
    p_{\perp}(1 + p_{\perp} + 2p_{\perp}^{\prime})
    \Big{]}
    \Big{)}}
    \\&\approx
    \Big{(}
    1 + \frac{p_{\perp}^{\prime}p_{\parallel} + Np_{\perp}^{\prime}p_{\perp}}{4p_{\parallel}}
    \Big{)}
    \Big{(}
    1
    -
    \frac{N}{2p_{\parallel}}
    \Big{[}
    p_{\parallel}(p_{\perp} + p_{\perp}^{\prime})
    +
    \\&+
    p_{\parallel}^{\prime}p_{\perp}
    +
    p_{\perp}(1 + p_{\perp} + 2p_{\perp}^{\prime})
    \Big{]}
    \Big{)}
    \\&\approx
    1
    -
    N\frac{p_{\perp}(1+p_{\parallel}) +p_{\parallel}^{\prime}p_{\perp} + p_{\parallel}p_{\perp}^{\prime}}{2p_{\parallel}}
    +
    \frac{p_{\perp}^{\prime}}{4}.
\end{aligned}
\end{equation}
As one can see from Fig.~\ref{fig:7}, the first-order approximation is accurate for few-photon GHZ states with high fidelity.

Next, we consider the typical experimental situation, where the collection efficiency is low and no frequency filtering is applied, which corresponds to $\eta_3 = \eta_2 = \eta$. Substituting~\eqref{eq:branching_coefficients_with_filters},\eqref{eq:branching_coefficients_with_filters_2} into Eq.~\eqref{eq:GHZ_fidelity_branching_conditional} yields
\begin{equation}
\begin{aligned}
\label{eq:GHZ_fidelity_branching_conditional_with_freq_filters}
    \tilde{\mathcal{F}}^{(N)}_{\mathrm{br,approx}}[\mathrm{GHZ}]
    \approx
    1
    -
    N(\frac{3\beta_{\perp} + \beta_{\perp}^{\prime}}{2})
    +
    \frac{\beta_{\perp}^{\prime}}{4}.
\end{aligned}
\end{equation}

Frequency filters can be applied to suppress the contribution from the undesired diagonal transition, which corresponds to the creation operator $\hat{B}^{\dagger}$ in \eqref{eq:freq_filter}. Filtering the undesired photons can be accounted for by putting $\eta_3 \ll \eta_2 = \eta$, which then yields
\begin{equation}
\begin{aligned}
\label{eq:GHZ_fidelity_branching_conditional_with_freq_filters_B}
    \tilde{\mathcal{F}}^{(N)}_{\mathrm{br,approx}}[\mathrm{GHZ}]
    &\approx
    1
    -
    N\frac{\beta_{\perp} + \beta_{\perp}^{\prime}}{2}
    +
    \frac{\beta_{\perp} + \beta_{\perp}^{\prime}}{4}
    \\&=
    1 - \frac{1}{2(B+1)}
    \Big{(} N - \frac{1}{2} \Big{)},
\end{aligned}
\end{equation}
where $B = (\beta_{\parallel} + \beta_{\parallel}^{\prime})/(\beta_{\perp} + \beta_{\perp}^{\prime})$ is the branching ratio between vertical and diagonal transitions in Fig.~\ref{fig:2}(d). As evident from Fig.~\ref{fig:7}, application of spectral filters improves the fidelity of the generated GHZ state with imperfect decay, which contrasts with the case of imperfect excitation(see Fig.~   \ref{fig:6}).

\subsubsection{Cluster state with branching errors}\label{sec:cluster_with_branching} 

The calculation of the cluster-state fidelity with branching errors is similar to that of the GHZ state.
Keeping only the terms that generate a correct cluster state, the single-round operator reads
\begin{equation}
    \begin{aligned}
    \label{eq:cluster_single_round_branching}
    \hat{O}_j^{\dagger}
    &=
    \ket{+}\bra{1}
    \Big{(}
    \sqrt{p_{\parallel}}
    \hat{A}^{\dagger}_{e,j}
    +
    \sqrt{p_{\perp}^{\prime}p_{\perp}}
    \hat{A}^{\dagger}_{e^{\prime},j}
    \hat{L}_j^{\dagger}
    \Big{)}
    \\&+
    \ket{-}\bra{0}
    \sqrt{p_{\parallel}}
    \hat{A}^{\dagger}_{l,j}
    +
    \ket{-}\bra{1}
    \sqrt{p_{\perp}^{\prime}p_{\parallel}}
    \hat{A}^{\dagger}_{l,j}
    \hat{E}_j^{\dagger}\delta_{j,1}.
    \end{aligned}
\end{equation}
The first two processes correspond to the ideal operation of the protocol, while the other two produce the correct state due to the incorrect operation, i.e. via the diagonal transition. 
Furthermore, the Kronecker-delta in the last term expresses the fact that only the first photon can produce the correct state when emitted in such process. Any other photon emitted in such process will results in a wrong spin-photon entangled state and hence will not contribute to the fidelity of the final state. As shown in Appendix.~\ref{subsec:appendix_D_Cluster}, the unconditional fidelity of the cluster state reads
\begin{equation}
    \begin{aligned}
    \label{eq:fidelity_of_cluster_branching_unconditional}
    \mathcal{F}^{(N)}_{\mathrm{br}}[\mathrm{Cl}]
    =
    \Big{(}
    p_{\parallel}
    +
    \frac{p_{\perp}p_{\perp}'}{4}
    \Big{)}^{N-1}
    \Big{(}
    p_{\parallel}
    +
    \frac{p_{\perp}p_{\perp}'}{4}
    +
    \frac{p_{\parallel}p_{\perp}'}{4}
    \Big{)}.
    \end{aligned}
\end{equation}
The success probability $P(n_1>0,..n_{N}>0)$~\eqref{eq:branching_success_probability_N} is calculated analogously to the GHZ state with the matrix elements
\begin{equation}
\begin{aligned}
    \nonumber
    M_{11} &=  M_{01} = 
    \frac{p_{\parallel}+p_{\perp}^2 + p_{\perp}p_{\parallel} + p_{\perp}^{\prime}p_{\parallel}+p_{\perp}p_{\parallel}^{\prime} + 2p_{\perp}p_{\perp}^{\prime}}{2},
    \\
    M_{10} &= M_{00} = \frac{1}{2}(p_{\parallel}+p_{\perp}).
\end{aligned}
\end{equation}
Normalizing the fidelity~\eqref{eq:fidelity_of_cluster_branching_unconditional} to the success probability yields the conditional fidelity shown in Fig.~\ref{fig:7}. In the first-order approximation, the fidelities of the cluster and GHZ state are identical and read
\begin{equation}
\label{eq:cluster_fidelity_branching_conditional}
    \boxed{
    \begin{aligned}
\tilde{\mathcal{F}}^{(N)}_{\mathrm{br,approx}}[\mathrm{Cl}]
    &=
    \tilde{\mathcal{F}}^{(N)}_{\mathrm{br,approx}}[\mathrm{GHZ}]
    \\&\approx
    1
    -
    N(\frac{3\beta_{\perp}+\beta_{\perp}^{\prime}}{2})
    +\frac{\beta_{\perp}^{\prime}}{4}
    \end{aligned}
    }
\end{equation}
and
\begin{equation}
\label{eq:Cluster_fidelity_branching_conditional_with_freq_filters}
\boxed{
\begin{aligned}
    \tilde{\mathcal{F}}^{(N)}_{\mathrm{br,approx}}[\mathrm{Cl}]
    &=
    \tilde{\mathcal{F}}^{(N)}_{\mathrm{br,approx}}[\mathrm{GHZ}]
    \\&\approx
    1
    -
    \Big{(}N - \frac{1}{2}\Big{)}
    \Big{(}\frac{\beta_{\perp} + \beta_{\perp}^{\prime}}{2}\Big{)},
\end{aligned}}
\end{equation}
without and with the frequency filters, respectively. The difference of $\beta_{\perp}$ between the frequency-filtered and unfiltered output states can be understood by considering the difference  between the effect of wrong photons emitted in the early and late time bins. With filtering, a diagonal transition in the late time bin does not create any photon and is thus removed by post selection. Bad effects thus only appear if the diagonal decay happens in the early time bin, followed by vertical decay  in the late time bin. With filtering the wrong decay thus only enters with half the probability $(\beta_{\perp}+\beta_{\perp}^{\prime})/2$ (except if if happens in the first round where it only disturbs coherences and thus have half the effect; hence the factor of $N-1/2)$. Without filtering, however, also diagonal decays going into the waveguide in the late time bin will be accepted and early emissions on the diagonal transition will have twice the probability to be accepted since two photons are emitted in this case. Both of these effects result  in the addition  of  the probability $\beta_{\perp}/2$.


\subsection{Total fidelity and nature of the errors}\label{sec:total_fidelity}
\begin{figure}[t]	
\includegraphics[width=0.9\columnwidth]{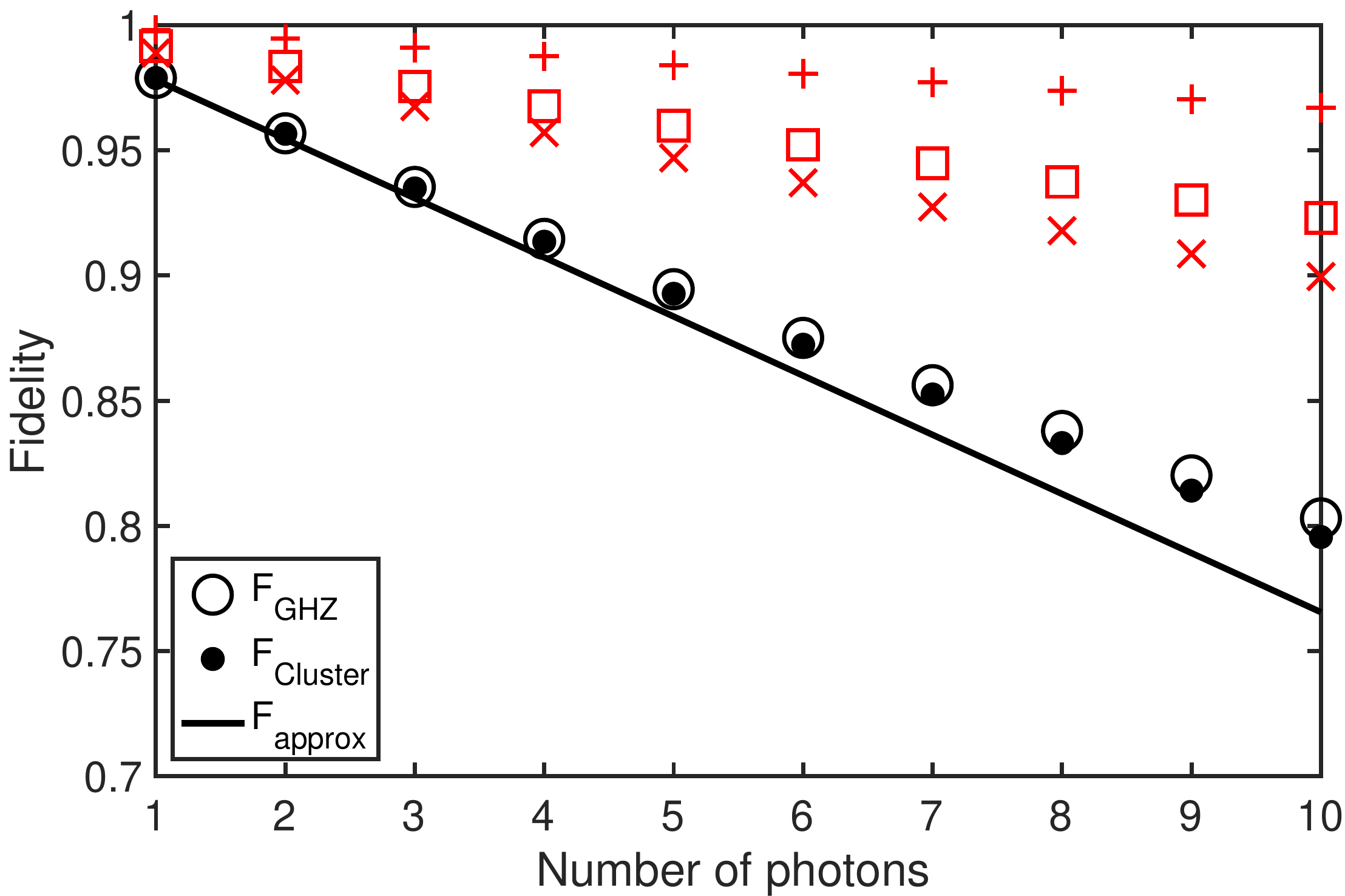}
\caption{\label{fig:8} Total fidelity versus number of photons with all imperfections and frequency filtering taken into account. Circles and dots show the fidelity of the GHZ and cluster states, respectively, and the solid line shows the first-order perturbative approximation of both states~\eqref{eq:final_fidelity_approx}. Red symbols correspond to the different contributions in Eq.~\eqref{eq:GHZ_final_fidelity}, namely, dephasing $\mathcal{F}_{\mathrm{ph}}$~(\textcolor{red}{$\times$}), excitation errors $\tilde{\mathcal{F}}_{\mathrm{exc}}$~(\textcolor{red}{$\Box$}), and imperfect branching $\tilde{\mathcal{F}}_{\mathrm{br}}$~(\textcolor{red}{$+$}). Here $\Delta = 2\pi\times 64$~GHz, $\gamma_{\mathrm{d}} = 0.06$~$\mathrm{ns}^{-1}$, $\gamma_{\mathrm{opt}} = 5.3$~$\mathrm{ns}^{-1}$, and $\beta$-factors are the same as in Fig.~\ref{fig:7}.}
\end{figure}

Taking into account all error sources discussed above, we approximate the fidelity of the GHZ and cluster states by the product of the individual fidelities, 
\begin{equation}
    \begin{aligned}
    \label{eq:GHZ_final_fidelity}
        \mathcal{F}_{\mathrm{GHZ/Cl}}
        =
        \mathcal{F}_{\mathrm{ph}}
        \times
        \tilde{\mathcal{F}}_{\mathrm{exc}}
        \times
        \tilde{\mathcal{F}}_{\mathrm{br}}.
        \end{aligned}
\end{equation}
Combining together Eqs.~(\ref{eq:fidelity_with_phonons},\ref{eq:excitation_with_filters_1st_order_text},\ref{eq:Cluster_fidelity_branching_conditional_with_freq_filters}), the first-order analytical expression is equivalent for both states and reads
\begin{equation}
    \begin{aligned}
    \label{eq:final_fidelity_approx}
        \mathcal{F}_{\mathrm{approx}}
        &=
        1
        +
        \frac{1}{4(B+1)}
        \\&-
        N\Big{(}
        \frac{\gamma_{\mathrm{d}}}{\gamma + 2\gamma_{\mathrm{d}}} 
        + 
        \frac{1}{2(B+1)}
        + 
        \frac{\sqrt{3}\pi}{8}\frac{\gamma}{\Delta}\Big{)}.
    \end{aligned}
\end{equation}
Figure~\ref{fig:8} shows the comparison between the full states fidelities and the first-order perturbative formula~\eqref{eq:final_fidelity_approx}.

So far we have only considered the fidelity of the state, which gives a simple characterization of the quality of the state. The fidelity, however, reduces all imperfections to a single number and  does not provide a full characterization of the complex many body state. More insight into the nature of the generated state can be obtained by further characterizing the nature of the errors occurring in the   generation        process.  

The error arising from dephasing affects the coherence between the internal internal states $\ket{0}$ and $\ket{1}$ or equivalently  the early and late time bins. This error can thus be characterised as a phase flip error  acting on the spin qubit before the last rotation pulse $\hat{R}$, or on the emitted photon. The excitation error takes a similar form. In the limit where we have efficient filtering of off-resonant photons, the excitation errors conserve the logical basis states. This error can  thus also be represented as the same phase flip.

The branching error is more complicated.  This process simultaneously affect the spin and the emitted photon. We analyse this situation more carefully in Appendix \ref{sec:appendix_E}, where we show that the dominant term from the branching error simultaneously affect two neighboring qubits.

A detailed understanding of nature of the error can be important for understanding potential applications of the states towards more advanced applications. As a particular example Refs.~\cite{Raussendorf_2007,PhysRevLett.98.190504,RAUSSENDORF20062242} investigate the application of cluster states for universal quantum computation. Here different error thresholds are derived for models where single qubit errors are applied after the state preparation and for more detailed model that include initialization, entanglement, memory, and measurement errors. In the former model an  error threshold of 3.2\%(1.4\%) per qubit are derived when computations are made with 3D(2D) cluster states. The infidelity per photon that we derive for the realistic parameters of Fig.~\ref{fig:8} is 2.1\%, which consists of 1.8\% of single-qubit errors and 0.3\% of two-qubit errors. The resulting error is thus roughly comparable to those models and it is very encouraging that our estimates for current experimental parameters are of a similar magnitude as fault-tolerance requirements. Extending cluster states to two or three dimensions will of course introduce additional errors that are likely to reduce the quality of the produced states below the requirements for fault-tolerance. Furthermore errors due to photon loss will also have to be accounted for. The exhaustive theoretical analysis conducted in this work, however,  identifies the main bottlenecks and provides a clear pathway for improving further beyond what is currently possible in the experiments. A full assessment of higher-dimensional states is outside of the scope of this work and calls for further extensive theoretical investigation. 

Finally, while  fault-tolerant quantum computation is beyond reach of any currently available technology, the requirements for quantum communication tasks are typically much less stringent. The generated states are time bin entangled states of photons and thus ideally suited for quantum communication through optical fibers. Indeed the generated GHZ-states can be directly applied to 
anonymous transmission~\cite{PhysRevA.98.052320,ASIACRYPT2005,ASIACRYPT2007}, secret sharing~\cite{PhysRevA.59.1829}, and leader election~\cite{7935fdc5be87497bb613fe4bbb79a6ad}. 
Taking the anonymous transmission protocol~\cite{PhysRevA.98.052320,ASIACRYPT2005,ASIACRYPT2007} as an example, the error threshold is known to depends on the number of communicating parties. 
According to the security analysis of Ref.~\cite{PhysRevA.98.052320}, the predicted error rates are within the threshold for up to at least fifty parties and almost an order of magnitude below the threshold for four parties.
 
\section{Conclusion}\label{sec:conclusion}
In conclusion, we have developed a theoretical approach for assessing the fidelity of entangled photonic states produced by a single quantum emitter. We derive simple analytical expressions for evaluating the fidelity of the generated states. These expression  provide a clear recipe for optimization of experimental parameters, such as photon emission rate and duration of the driving laser pulses. Our framework can be straightforwardly applied to a broad range of quantum emitters, including semiconductor quantum dots coupled to nanophotonic structures, defect centers in solids, and atoms in cavities. With the rapid experimental developments in quantum nanophotonics we expect that these results can form the basis of  near-future realisations of multiphoton emitters with a performance exceeding existing methods.

As we discuss here and in the companion paper~\cite{Tiurev2020b}, the considered time-bin generation protocol appears to be a particularly promising approach for the sequential production of entangled photons from quantum dot emitters.  Here it is highly appealing  that our analysis shows that the output state is insensitive to a number of slow drifts of experimental parameters. Therefore, for instance the very short $T_2^*$ coherence time of spin qubits in quantum dots, which is a limiting factor in many quantum-information applications, does not compromise the protocol considered here.  Based on our theoretical considerations, we predict that quantum dot emitters currently available  can be used to produce five-photon GHZ and cluster states with fidelities of approximately 80\%. A fidelity above the 50\% level is present in  states containing up to 10 subsequent photons. This is comparable to the state of the art achieved thus far with other methods~\cite{PhysRevLett.117.210502,PhysRevLett.121.250505}, but the generation rate is expected to be much higher with the presented deterministic approach.

\begin{acknowledgments}
We gratefully acknowledge financial support from Danmarks Grundforskningsfond~(DNRF 139, Hy-Q Center for Hybrid Quantum Networks), the European Research Council (ERC Advanced Grant `SCALE'), and the European Union Horizon 2020 research and innovation programme under grant agreement N\textsuperscript{\underline{o}}~820445 and project name Quantum Internet Alliance.
\end{acknowledgments}

\clearpage
\onecolumngrid
\appendix

\section{Ideal scheme for generation of the cluster state} \label{sec:Appendix_0}

Below we prove that the ideal scheme generates the cluster state for arbitrary large number of photons. Consider a  single round of the protocol discussed in Sec.~\ref{sec:protocol}, which can be written as
\begin{equation}\label{eq:O_output}
    \begin{aligned}
        \hat{O}^{\dagger}
        &=
        \hat{H}\hat{L}^{\dagger}\hat{X}\hat{E}^{\dagger}
        =
        \frac{1}{\sqrt{2}}
        \Big{(}
        (\ket{0}+\ket{1})\bra{0}
        +
        (\ket{0}-\ket{1})\bra{1}
        \Big{)}
        \Big{(}
        \ket{0}\bra{0}
        +
        \ket{1}\bra{1}\hat{a}^{\dagger}_l
        \Big{)}
        \Big{(}
        \ket{0}\bra{1}
        +
        \ket{1}\bra{0}
        \Big{)}
        \Big{(}
        \ket{0}\bra{0}
        +
        \ket{1}\bra{1}\hat{a}^{\dagger}_e
        \Big{)}
        \\&=
        \frac{1}{\sqrt{2}}
        \Big{(}\ket{0}+\ket{1}\Big{)}\bra{1}\hat{a}^{\dagger}_e
        +
        \frac{1}{\sqrt{2}}
        \Big{(}\ket{0}-\ket{1}\Big{)}\bra{0}\hat{a}^{\dagger}_l,
    \end{aligned}
\end{equation}
where $\hat{E}^{\dagger}$ and $\hat{L}^{\dagger}$ are, respectively, the operators corresponding to the generations of an early and a late photons. For convenience, let us change basis and choose the logical spin states as
$\ket{0} \rightarrow \ket{1}$, $\ket{1} \rightarrow \ket{0}$ and the logical photon states as $\hat{a}^{\dagger}_e \rightarrow \hat{a}^{\dagger}_0$, $\hat{a}^{\dagger}_l \rightarrow -\hat{a}^{\dagger}_1$, thus turning the operator of Eq.~\eqref{eq:O_output} into
\begin{equation}\label{eq:O_good}
    \hat{O}^{\dagger} = \ket{+}\bra{0}\hat{a}^{\dagger}_0 + \ket{-}\bra{1}\hat{a}^{\dagger}_1,
\end{equation}
where $\hat{a}_0^{\dagger}$ and $\hat{a}_1^{\dagger}$ create photons in states $\ket{0}$ and $\ket{1}$, respectively, and $\ket{\pm} = (\ket{0} \pm \ket{1})/\sqrt{2}$. By definition, an ($N$+1)-qubit cluster state is a simultaneous eigenstate of the operators $\hat{g}_i$, where 
\begin{equation}
    \begin{aligned}
        \hat{g}_i &= \hat{Z}_{i-1}\hat{X}_{i}\hat{Z}_{i+1} \: \forall i=[1,N-1]
        \\
        \hat{g}_0 &= \hat{X}_{0}\hat{Z}_{1}
        \\
        \hat{g}_N &= \hat{Z}_{N-1}\hat{X}_{N},
    \end{aligned}
\end{equation}
with $\hat{Z}$ and $\hat{X}$ being the Pauli-Z and Pauli-X matrices, respectively. We will now prove the following theorem:

\textit{\textbf{Theorem.}} Assume that $\ket{\Psi_N}$ is a cluster state generated by the action of the operator $\hat{O}^{\dagger}$~\eqref{eq:O_good}, such that $g_i = 1$ $\forall$ $i = [0,N]$. Then the state $\ket{\Psi_{N+1}} = \hat{O}^{\dagger}\ket{\Psi_N}$ is also a cluster state with $g_i = 1$ $\forall$ $i = [0,N+1]$.

\textit{\textbf{Proof.}}
We start by writing the operator $\hat{O}^{\dagger}$ from~\eqref{eq:O_good} in two bases,
\begin{equation}
    \label{eq:O_x_z}
    \hat{O}^{\dagger}_{x \leftarrow z}
    =
    \ket{+}\hat{a}^{\dagger}_0 \bra{0} 
    + 
    \ket{-}\hat{a}^{\dagger}_1 \bra{1}
\end{equation}
and
\begin{equation}    
    \label{eq:O_z_x}
    \hat{O}^{\dagger}_{z \leftarrow x}
    =
    \frac{1}{\sqrt{2}}
    \Big{(}
    \ket{0}\hat{a}^{\dagger}_+ + \ket{1}\hat{a}^{\dagger}_-
    \Big{)}
    \bra{+}
    +
    \frac{1}{\sqrt{2}}
    \Big{(}
    \ket{0}\hat{a}^{\dagger}_- + \ket{1}\hat{a}^{\dagger}_+
    \Big{)}
    \bra{-},
\end{equation}
where $\hat{a}_{\pm} = (\hat{a}_0 \pm \hat{a}_1)/\sqrt{2}$.

Since the cluster state is an eigenstate of $\hat{g}_N$ and $\hat{g}_{N-1}$, the general form of the state $\ket{\Psi_N}$ can be written in each of the two bases,
\begin{equation}
    \label{eq:psi_xz}    
    \ket{\Psi_N^{(xz)}} = \ket{+0}\ket{\Psi_{N-2}^{(+0)}} + \ket{-1}\ket{\Psi_{N-2}^{(-1)}}
    \end{equation}
and
\begin{equation}
    \label{eq:psi_zx}
    \ket{\Psi_N^{(zx)}} = 
    \ket{0+0}\ket{\Psi_{N-3}^{(0+0)}}
    +
    \ket{1-0}\ket{\Psi_{N-3}^{(1-0)}}
    +
    \ket{0-1}\ket{\Psi_{N-3}^{(0-1)}}
    +
    \ket{1+1}\ket{\Psi_{N-3}^{(1+1)}},
\end{equation}
where we label the spin-photon states such that the spin state is always the ket-vector furthest to the left followed by the photon states, i.e. $\ket{\Psi_N} = \ket{\mathrm{Spin},\mathrm{Photon}_N,\mathrm{Photon}_{N-1},..}$.

To prove that the state $\ket{\Psi_{N+1}}$ is a cluster state, we need to show that all stabilizers obey $g_i = 1$ $\forall$ $i=[0,N+1]$. The operator $\hat{O}^{\dagger}$ only acts on the qubit {\#} N and adds the qubit {\#} (N+1). Thus, it does not change the value of the stabilizers $g_1$ to $g_{N-2}$ and it suffices to prove that the eigenvalues of the stabilizers $\hat{g}_{N-1}$, $\hat{g}_{N}$, and $\hat{g}_{N+1}$ are equal to 1.
First, we act with the operator $\hat{O}^{\dagger}_{x \leftarrow z}$~\eqref{eq:O_x_z} on the state $\ket{\Psi_N^{(zx)}}$~\eqref{eq:psi_zx},
\begin{equation}
    \begin{aligned}
    \ket{\Psi_{N+1}^{xz}} 
    &=
    \hat{O}^{\dagger}_{x \leftarrow z} \ket{\Psi_N^{(zx)}}
    \\&=
    \ket{+0+0}\ket{\Psi_{N-3}^{(0+0)}}
    +
    \ket{-1-0}\ket{\Psi_{N-3}^{(1-0)}}
    +
    \ket{+0-1}\ket{\Psi_{N-3}^{(0-1)}}
    +
    \ket{-1+1}\ket{\Psi_{N-3}^{(1+1)}}.
    \end{aligned}
\end{equation}
From the second line of the equation above it follows that the $g_{N+1}=1$ and  $g_{N-1}=1$. 

Next, we act with the operator $\hat{O}^{\dagger}_{z \leftarrow x}$~\eqref{eq:O_z_x} on the state $\ket{\Psi_N^{(xz)}}$~\eqref{eq:psi_xz}, 
\begin{equation}
    \begin{aligned}
    \ket{\Psi_{N+1}^{zx}} 
    &=
    \hat{O}^{\dagger}_{z \leftarrow x} \ket{\Psi_N^{(xz)}}
    \\&=
    \ket{0+0}\ket{\Psi_{N-2}^{+0}}
    +
    \ket{1-0}\ket{\Psi_{N-2}^{-0}}
    +
    \ket{0-1}\ket{\Psi_{N-2}^{-1}}
    +
    \ket{1+1}\ket{\Psi_{N-2}^{+1}}
    \end{aligned}
\end{equation}
Therefore, the state $\ket{\Psi_{N+1}^{zx}}$ obeys $g_N = 1$ and we have proven that all stabilizers obey $g_i = 1$. Thus, an operator $\hat{O}^{\dagger}$ takes an N-qubit cluster state to an (N+1)-qubit cluster state. This concludes the proof of the theorem. $\blacksquare$

To complete the proof that the procedure creates a cluster state we still need to show that we can generate a cluster state for a small $N$. 
This can be proven by applying the operator $\hat{O}^{\dagger}$ of Eq.~\eqref{eq:O_good} twice to a qubit initially prepared in $\ket{\Psi_0}=\ket{+}$, which produces a  state
\begin{equation}
    \ket{\Psi_2}
    =
\hat{O}^{\dagger}_2\hat{O}^{\dagger}_1\ket{\Psi_0,\emptyset}
    =
    \frac{1}{\sqrt{2}}
    \Big{(}\ket{+0+} + \ket{-1-}\Big{)}
    =
    \frac{1}{2}
    \Big{(}
    \ket{0+0} + \ket{1+1} + \ket{0-1} + \ket{1-0}
    \Big{)}.
\end{equation}
This state can be directly verified to be a cluster state.

\section{Fidelities of non-spin-mixing errors}\label{sec:Appendix_A}
\emph{GHZ state} --- The $N$-photon operators that enter the expression for the operational fidelity~\eqref{eq:exp_fidelity} read
\begin{equation}
    \begin{aligned}
    \label{eq:GHZ_operators_appendix}
    \hat{o}_1
    ..
    \hat{o}_N
    \hat{O}^{\dagger}_N
    ..
    \hat{O}^{\dagger}_1
    &=
    \ket{1}
    \bra{1}
    \hat{a}_{e,1}(t_1)
    \hat{A}^{\dagger}_{e,1}
    ..
    \hat{a}_{e,N}(t_N)
    \hat{A}^{\dagger}_{e,N}
    +
    \ket{0}
    \bra{0}
    \hat{a}_{l,1}(t_1)
    \hat{A}^{\dagger}_{l,1}
    ..
    \hat{a}_{l,N}(t_N)
    \hat{A}^{\dagger}_{l,N}.
    \end{aligned}
    \end{equation}
Inserting~\eqref{eq:GHZ_operators_appendix} and its Hermitian conjugate into equation for the operational fidelity~\eqref{eq:exp_fidelity}, we obtain
\begin{equation}
    \begin{aligned}
    \label{eq:GHZ_exp_fidelity_appendix}
    &\mathcal{F}^{(N)}_{}[\mathrm{GHZ}]
    \\&=
    \mathrm{Tr}_{\mathrm{env}}
    \Big{\{}
    \int_0^{\infty}
    \mathrm{d} t_N
    ..
    \int_0^{\infty}
    \mathrm{d} t_1
    \bra{\emptyset}
    \bra{\Psi_0}
    \Big{(}
    \ket{1}
    \bra{1}
    \hat{a}_{e,1}(t_1)
    \hat{A}^{\dagger}_{e,1}
    ..
    \hat{a}_{e,N}(t_N)
    \hat{A}^{\dagger}_{e,N}
    +
    \ket{0}
    \bra{0}
    \hat{a}_{l,1}(t_1)
    \hat{A}^{\dagger}_{l,1}
    ..
    \hat{a}_{l,N}(t_N)
    \hat{A}^{\dagger}_{l,N}
    \Big{)}
    \ket{\Psi_0}
    \ket{\emptyset}
    \\&
    \bra{\emptyset}
    \bra{\Psi_0}
    \Big{(}
    \ket{1}
    \bra{1}
    \hat{A}_{e,1}
    \hat{a}^{\dagger}_{e,1}(t_1)
    ..
    \hat{A}_{e,N}
    \hat{a}^{\dagger}_{e,N}(t_N)
    +
    \ket{0}
    \bra{0}
    \hat{A}_{l,1}
    \hat{a}^{\dagger}_{l,1}(t_1)
    ..
    \hat{A}_{l,N}
    \hat{a}^{\dagger}_{l,N}(t_N)
    \Big{)}
    \ket{\Psi_0}
    \ket{\emptyset}
    \\&=
    \frac{1}{4}
    \mathrm{Tr}_{\mathrm{env}}
    \Big{\{}
    \int_0^{\infty}
    \mathrm{d} t_N
    ..
    \int_0^{\infty}
    \mathrm{d} t_1
    \bra{\emptyset}
    \Big{(}
    \hat{a}_{e,1}(t_1)
    \hat{A}^{\dagger}_{e,1}
    ..
    \hat{a}_{e,N}(t_N)
    \hat{A}^{\dagger}_{e,N}
    +
    \hat{a}_{l,1}(t_1)
    \hat{A}^{\dagger}_{l,1}
    ..
    \hat{a}_{l,N}(t_N)
    \hat{A}^{\dagger}_{l,N}
    \Big{)}
    \ket{\emptyset}
    \\&
    \bra{\emptyset}
    \Big{(}
    \hat{A}_{e,1}
    \hat{a}^{\dagger}_{e,1}(t_1)
    ..
    \hat{A}_{e,N}
    \hat{a}^{\dagger}_{e,N}(t_N)
    +
    \hat{A}_{l,1}
    \hat{a}^{\dagger}_{l,1}(t_1)
    ..
    \hat{A}_{l,N}
    \hat{a}^{\dagger}_{l,N}(t_N)
    \Big{)}
    \ket{\emptyset}
    \Big{\}}
    \\&=
    \frac{1}{4}
    \mathrm{Tr}_{\mathrm{env}}
    \Big{\{}
    \int_0^{\infty}
    \mathrm{d} t_N
    ..
    \int_0^{\infty}
    \mathrm{d} t_1
    \bra{\emptyset}
    \Big{(}
    \hat{a}_{e,1}(t_1)
    \hat{A}^{\dagger}_{e,1}
    ..
    \hat{a}_{e,N}(t_N)
    \hat{A}^{\dagger}_{e,N}
    \ket{\emptyset}
    \bra{\emptyset}
    \hat{A}_{e,1}
    \hat{a}^{\dagger}_{e,1}(t_1)
    ..
    \hat{A}_{e,N}
    \hat{a}^{\dagger}_{e,N}(t_N)
    \Big{)}
    \ket{\emptyset}
    \\&+
    \int_0^{\infty}
    \mathrm{d} t_N
    ..
    \int_0^{\infty}
    \mathrm{d} t_1
    \bra{\emptyset}
    \Big{(}
    \hat{a}_{l,1}(t_1)
    \hat{A}^{\dagger}_{l,1}
    ..
    \hat{a}_{l,N}(t_N)
    \hat{A}^{\dagger}_{l,N}
    \ket{\emptyset}
    \bra{\emptyset}
    \hat{A}_{e,1}
    \hat{a}^{\dagger}_{e,1}(t_1)
    ..
    \hat{A}_{e,N}
    \hat{a}^{\dagger}_{e,N}(t_N)
    \Big{)}
    \ket{\emptyset}
    \\&+
    \int_0^{\infty}
    \mathrm{d} t_N
    ..
    \int_0^{\infty}
    \mathrm{d} t_1
    \bra{\emptyset}
    \Big{(}
    \hat{a}_{e,1}(t_1)
    \hat{A}^{\dagger}_{e,1}
    ..
    \hat{a}_{e,N}(t_N)
    \hat{A}^{\dagger}_{e,N}
    \ket{\emptyset}
    \bra{\emptyset}
    \hat{A}_{l,1}
    \hat{a}^{\dagger}_{l,1}(t_1)
    ..
    \hat{A}_{l,N}
    \hat{a}^{\dagger}_{l,N}(t_N)
    \Big{)}
    \ket{\emptyset}
    \\&+
    \int_0^{\infty}
    \mathrm{d} t_N
    ..
    \int_0^{\infty}
    \mathrm{d} t_1
    \bra{\emptyset}
    \Big{(}
    \hat{a}_{l,1}(t_1)
    \hat{A}^{\dagger}_{l,1}
    ..
    \hat{a}_{l,N}(t_N)
    \hat{A}^{\dagger}_{l,N}
    \ket{\emptyset}
    \bra{\emptyset}
    \hat{A}_{l,1}
    \hat{a}^{\dagger}_{l,1}(t_1)
    ..
    \hat{A}_{l,N}
    \hat{a}^{\dagger}_{l,N}(t_N)
    \Big{)}
    \ket{\emptyset}
    \Big{\}}
    \\&=
    \frac{1}{4}
    \mathrm{Tr}_{\mathrm{env}}
    \sum_{u,v=e,l}
    \Big{(}
    \int_0^{\infty}
    \mathrm{d} t
    \bra{\emptyset}
    \hat{a}_{u}(t)
    \hat{A}_{u}^{\dagger}
    \ket{\emptyset}
    \bra{\emptyset}
    \hat{A}_{v}
    \hat{a}_{v}^{\dagger}(t)
    \ket{\emptyset}
    \Big{)}^N.
    \end{aligned}
    \end{equation}
    
In the last step we have used that the photonic operators $\hat{A}_u$ for different time periods commute. This means that the photonic part of the matrix element can be separated  into products. Note, however, that the operators $\hat{A}^{\dagger}_{u}$ may contain couplings to different degrees of freedom, for which this factorization may not be the applicable, e.g., in Sec.~\ref{subsubsec:phonon_induced_pure dephasing}, $\hat{A}^{\dagger}_{u}$ contain the coupling to a phononic  environment. In this case the $N$th order product of $\hat{A}_u^{\dagger}$ operators should in principle  be evaluated as a suitable time ordered product for different periods. We will, however, only consider situations in which this product can be completely separated, e.g., a Markovian phononic reservoir.     
\\

In general the approximation applied here is reminiscent of the Markovian approximation often employed in quantum optics, but not exactly the same. In particular slowly varying classical parameters as considered in Sec.~\ref{subsubsec:ground_state_dephasing}, do not fit into the usual Markovian approximation, but is still compatible with~\eqref{eq:GHZ_exp_fidelity_appendix}, provided that the average over the classical parameter (implied by $\mathrm{Tr}_{\mathrm{env}}$) is performed for the final $N$th order product and not for each term individually. On the other hand, the situation would be more complicated if we were, e.g., considering a non-Markovian phononic reservoir.
\\

\emph{Cluster state} --- For the cluster state the two-photon operator is 
\begin{equation}
    \begin{aligned}
    &\hat{o}_1(t_1)\hat{o}_2(t_2)\hat{O}_2^{\dagger}\hat{O}_1^{\dagger}
    \\&=
    \Big{(}
    \ket{1}\bra{0}\hat{a}_{e,1}(t_{1}) + 
    \ket{0}\bra{1}\hat{a}_{l,1}(t_{1})
    \Big{)}
    \hat{H}^{\dagger}
    \Big{(}
    \ket{1}
    \bra{1}
    \hat{a}_{e,2}(t_{2})
    \hat{A}^{\dagger}_{e,2}
    +
    \ket{0}
    \bra{0}
    \hat{a}_{l,2}(t_{2})
    \hat{A}^{\dagger}_{l,2}
    \Big{)}
    \hat{H}
    \Big{(}
    \ket{0}\bra{1}\hat{A}^{\dagger}_{e,1} + \ket{1}\bra{0}\hat{A}^{\dagger}_{l,1}
    \Big{)}
    \\&=
    \frac{1}{2}
    \Big{[}
    \ket{1}\bra{1}
    \hat{a}_{e,1}(t_{1})
    \hat{A}_{e,1}^{\dagger}
    \otimes
    \Big{(}
    \hat{a}_{e,2}(t_{2})
    \hat{A}_{e,2}^{\dagger}
    +
    \hat{a}_{l,2}(t_{2})
    \hat{A}_{l,2}^{\dagger}
    \Big{)}
    +
    \ket{1}\bra{1}
    \hat{a}_{l,1}(t_{1})
    \hat{A}_{l,1}^{\dagger}
    \otimes
    \Big{(}
    \hat{a}_{e,2}(t_{2})
    \hat{A}_{e,2}^{\dagger}
    +
    \hat{a}_{l,2}(t_{2})
    \hat{A}_{l,2}^{\dagger}
    \Big{)}
    \Big{]}
    \\&=
    \frac{1}{2}
    \Big{(}
    \ket{1}\bra{1}
    \hat{a}_{e,1}(t_{1})
    \hat{A}_{e,1}^{\dagger}
    +
    \ket{0}\bra{0}
    \hat{a}_{l,1}(t_{1})
    \hat{A}_{l,1}^{\dagger}
    \Big{)}
    \otimes
    \Big{(}
    \hat{a}_{e,2}(t_{2})
    \hat{A}_{e,2}^{\dagger}
    +
    \hat{a}_{l,2}(t_{2})
    \hat{A}_{l,2}^{\dagger}
    \Big{)},
    \end{aligned}
\end{equation}
where we have omitted cross-terms such as 
$\hat{a}_{l,1}(t_{1})\hat{a}_{l,2}(t_{2})\hat{A}_{l,2}^{\dagger}\hat{A}_{e,1}^{\dagger}$ since they will vanish when sandwiched with the photon vacuum, 
$\bra{\emptyset_1} \hat{a}_{l,1}(t_{1})\hat{A}_{e,1}^{\dagger} \ket{\emptyset_1}= 0$.
For arbitrary $N$, this generalizes to
\begin{equation}
\begin{aligned}
    \hat{o}_1(t_1)..\hat{o}_N(t_N)
    \hat{O}_N^{\dagger}..\hat{O}_1^{\dagger}
    =
    \frac{1}{2^{N-1}}
    \Big{(}
    \ket{1}\bra{1}
    \hat{a}_{e,1}(t_{1})
    \hat{A}_{e,1}^{\dagger}
    +
    \ket{0}\bra{0}
    \hat{a}_{l,1}(t_{1})
    \hat{A}_{l,1}^{\dagger}
    \Big{)}
    \otimes
    \prod_{j=2}^{N}
    \Big{(}
    \hat{a}_{e,j}(t_{j})
    \hat{A}_{e,j}^{\dagger}
    +
    \hat{a}_{l,j}(t_{j})
    \hat{A}_{l,j}^{\dagger}
    \Big{)},
\end{aligned}    
\end{equation}
and applying it to the initial spin state 
$\ket{\Psi_0} = (\ket{0}+\ket{1})/\sqrt{2}$ we find
\begin{equation}
\begin{aligned}
\label{eq:operators_cluster_N_nonmix}
    \bra{\Psi_0}
    \hat{o}_1(t_1)..\hat{o}_N(t_N)
    \hat{O}_N^{\dagger}..\hat{O}_1^{\dagger}
    \ket{\Psi_0}
    =
    \frac{1}{2^{N}}
    \prod_{j=1}^{N}
    \Big{(}
    \hat{a}_{e,j}(t_{j})
    \hat{A}_{e,j}^{\dagger}
    +
    \hat{a}_{l,j}(t_{j})
    \hat{A}_{l,j}^{\dagger}
    \Big{)}.
\end{aligned}    
\end{equation}
Substituting~\eqref{eq:operators_cluster_N_nonmix} into~\eqref{eq:exp_fidelity}, one arrives at
\begin{equation}
\begin{aligned}
\label{eq:Cluster_exp_fidelity_appendix}
    &\mathcal{F}^{(N)}_{}[\mathrm{Cl}]
    \\&=
    \mathrm{Tr}_{\mathrm{env}}
    \int_0^{\infty}
    \mathrm{d} t_N
    ..
    \int_0^{\infty}
    \mathrm{d} t_1
    \bra{\emptyset}
    \bra{\Psi_0}
    \hat{o}_1(t_1)..\hat{o}_N(t_N)
    \hat{O}_N^{\dagger}..\hat{O}_1^{\dagger}
    \ket{\Psi_0}
    \ket{\emptyset}
    \bra{\emptyset}
    \bra{\Psi_0}
    \hat{O}_1..\hat{O}_N
    \hat{o}^{\dagger}_N(t_N)..\hat{o}^{\dagger}_1(t_1)
    \ket{\Psi_0}
    \ket{\emptyset}
    \\&=
    \frac{1}{2^{2N}}
    \mathrm{Tr}_{\mathrm{env}}
    \int_0^{\infty}
    \mathrm{d} t_N
    ..
    \int_0^{\infty}
    \mathrm{d} t_1
    \bra{\emptyset}
    \prod_{j=1}^{N}
    \Big{(}
    \hat{a}_{e,j}(t_{j})
    \hat{A}_{e,j}^{\dagger}
    +
    \hat{a}_{l,j}(t_{j})
    \hat{A}_{l,j}^{\dagger}
    \Big{)}
    \ket{\emptyset}
    \bra{\emptyset}
    \Big{(}
    \hat{A}_{e,j}
    \hat{a}_{e,j}^{\dagger}(t_{j})
    +
    \hat{A}_{l,j}
    \hat{a}_{l,j}^{\dagger}(t_{j})
    \Big{)}
    \ket{\emptyset}
    \\&=
    \frac{1}{2^{2N}}
    \mathrm{Tr}_{\mathrm{env}}
    \int_0^{\infty}
    \mathrm{d} t_1
    \bra{\emptyset_1}
    \Big{(}
    \hat{a}_{e,1}(t_{1})
    \hat{A}_{e,1}^{\dagger}
    +
    \hat{a}_{l,1}(t_{1})
    \hat{A}_{l,1}^{\dagger}
    \Big{)}
    \ket{\emptyset_1}
    \bra{\emptyset_1}
    \Big{(}
    \hat{A}_{e,1}
    \hat{a}_{e,1}^{\dagger}(t_{1})
    +
    \hat{A}_{l,1}
    \hat{a}_{l,1}^{\dagger}(t_{1})    
    \Big{)}
    \ket{\emptyset_1}
    \times
    ..
    \\&
    ..\times
    \int_0^{\infty}
    \mathrm{d} t_N
    \bra{\emptyset_N}
    \Big{(}
    \hat{a}_{e,N}(t_{N})
    \hat{A}_{e,N}^{\dagger}
    +
    \hat{a}_{l,N}(t_{N})
    \hat{A}_{l,N}^{\dagger}
    \Big{)}
    \ket{\emptyset_N}
    \bra{\emptyset_N}
    \Big{(}
    \hat{A}_{e,N}
    \hat{a}_{e,N}^{\dagger}(t_{N})
    +
    \hat{A}_{l,N}
    \hat{a}_{l,N}^{\dagger}(t_{N})    
    \Big{)}
    \ket{\emptyset_N}
    \\&=
    \mathrm{Tr}_{\mathrm{env}}
    \Big{(}
    \frac{1}{4}
    \sum_{u,v=e,l}
    \int_0^{\infty}
    \mathrm{d} t
    \bra{\emptyset}
    \hat{a}_{u}(t)
    \hat{A}_{u}^{\dagger}
    \ket{\emptyset}
    \bra{\emptyset}
    \hat{A}_{v}
    \hat{a}_{v}^{\dagger}(t)
    \ket{\emptyset}
    \Big{)}^N.
\end{aligned}    
\end{equation}
In the last step we have again applied the approximation discussed after Eq.~\eqref{eq:GHZ_exp_fidelity_appendix}. 

\section{Two-photon emissions}\label{sec:Appendix_B}
We are interested in taking into account the action of the temporal and frequency filters on the output state~\eqref{eq:two_photon_state}.
\\

\emph{Frequency filters} ---
For convenience, we start from frequency filtering. Applying the transformation in Eq.~\eqref{eq:freq_filter}, the creation operators~\eqref{eq:Q_operator} become
\begin{equation}
    \begin{aligned}
    \label{eq:Q_operator_freq_filtered}
    \hat{\tilde{Q}}^{\dagger}_{u,j} &= 
    \Big{(}
    c_1 + 
    c_2\sqrt{\eta_2}\hat{A}_0^{\dagger,u} +
    c_2(1-\sqrt{\eta_2})\hat{\tilde{A}}_0^{\dagger,u}
    \Phi_1\hat{A}_{p_1}^{\dagger,u}+
    \Phi_2\sqrt{\eta_2}\hat{A}_{p_2}^{\dagger,u}\hat{A}_{0}^{\dagger,u}+
    \Phi_2\sqrt{1-\eta_2}\hat{A}_{p_2}^{\dagger,u}\hat{\tilde{A}}_{0}^{\dagger,u}
    \Big{)}_j
    \\&\times
    \Big{(}
    c_1 + 
    c_2\sqrt{\eta_3}\hat{B}_0^{\dagger,v} +
    c_2(1-\sqrt{\eta_3})\hat{\tilde{B}}_0^{\dagger,v}
    \Phi_1\hat{A}_{p_1}^{\dagger,v}+
    \Phi_2\sqrt{\eta_3}\hat{B}_{p_2}^{\dagger,v}\hat{B}_{0}^{\dagger,v}+
    \Phi_2\sqrt{1-\eta_3}\hat{B}_{p_2}^{\dagger,v}\hat{\tilde{B}}_{0}^{\dagger,v}
    \Big{)}_j,
    \end{aligned}
\end{equation}
where $j$ is the photon number and $\{u,v\} = \{e,l\}$, $u \neq v$.

\emph{Temporal filters} --- 
We condition on detecting a photon in the decay period of either the  early or the late pulse. We thus apply the projector $\hat{P}_{n_{0j} > 0}$ on the operator above thus keeping only the terms that correspond to receiving at least one photon after each excitation pulse,
\begin{equation}
    \begin{aligned}
    \label{eq:Q_operator_fully_filtered_appendix}
    \hat{P}_{n_{0j} > 0}
    \hat{\tilde{Q}}^{\dagger}_{u,j} 
    &= 
    c_1c_3 \sqrt{\eta_3}\hat{B}_0^{\dagger,v}
    +
    c_1\Phi_3 \sqrt{\eta_3}\hat{B}_0^{\dagger,v}\hat{B}_{p_2}^{\dagger,v}
    +
    c_2c_0 \sqrt{\eta_2}\hat{A}_0^{\dagger,u}
    +
    c_2\Phi_3 \sqrt{\eta_2}\sqrt{\eta_3}\hat{A}_0^{\dagger,u}\hat{B}_{0}^{\dagger,v}
    +
    c_2c_3 \sqrt{\eta_2}\sqrt{1-\eta_3}\hat{A}_0^{\dagger,u}\hat{\tilde{B}}_{0}^{\dagger,v}
    \\&+
    c_2\Phi_0 \sqrt{\eta_2}\hat{A}_0^{\dagger,u}\hat{B}_{p_1}^{\dagger,v}
    +
    c_2\Phi_3 \sqrt{\eta_2}\sqrt{\eta_3}\hat{A}_0^{\dagger,u}\hat{B}_{p_2}^{\dagger,v}\hat{B}_{0}^{\dagger,v}
    +
    c_2\Phi_3 \sqrt{\eta_2}\sqrt{1-\eta_3}\hat{A}_0^{\dagger,u}\hat{B}_{p_2}^{\dagger,v}\hat{\tilde{B}}_{0}^{\dagger,v}
    +
    c_2c_3 \sqrt{\eta_3}\sqrt{1-\eta_2}\hat{\tilde{A}}_0^{\dagger,u}\hat{\tilde{B}}_{0}^{\dagger,v}
    \\&+
    c_2\Phi_3 \sqrt{\eta_3}\sqrt{1-\eta_2}\hat{\tilde{A}}_0^{\dagger,u}\hat{B}_{p_2}^{\dagger,v}\hat{B}_{0}^{\dagger,v}
    +
    c_3\Phi_1 \sqrt{\eta_3}\hat{\tilde{A}}_{p_1}^{\dagger,u}\hat{B}_{0}^{\dagger,v}
    +
    \Phi_1\Phi_3 \sqrt{\eta_3}\hat{\tilde{A}}_{p_1}^{\dagger,u}\hat{B}_{p_2}^{\dagger,v}\hat{B}_{0}^{\dagger,v}
    +
    \Phi_2 c_0 \sqrt{\eta_2}\hat{\tilde{A}}_{p_2}^{\dagger,u}\hat{A}_{0}^{\dagger,u}
    \\&+
    \Phi_2 c_3 \sqrt{\eta_2}\sqrt{\eta_3}\hat{\tilde{A}}_{p_2}^{\dagger,u}\hat{A}_{0}^{\dagger,u}\hat{B}_{0}^{\dagger,v}
    +
    \Phi_2 c_3 \sqrt{\eta_2}\sqrt{1-\eta_3}\hat{\tilde{A}}_{p_2}^{\dagger,u}\hat{A}_{0}^{\dagger,u}\hat{\tilde{B}}_{0}^{\dagger,v}
    +
    \Phi_2 \Phi_0 \sqrt{\eta_2}
    \hat{\tilde{A}}_{p_2}^{\dagger,u}\hat{A}_{0}^{\dagger,u}\hat{\tilde{B}}_{0}^{\dagger,v}
    \\&+
    \Phi_2 \Phi_3 \sqrt{\eta_2}\sqrt{\eta_3}\hat{\tilde{A}}_{p_2}^{\dagger,u}\hat{A}_{0}^{\dagger,u}\hat{B}_{0}^{\dagger,v}\hat{B}_{p_2}^{\dagger,v}
    +
    \Phi_2 \Phi_3 \sqrt{\eta_2}\sqrt{1-\eta_3}\hat{\tilde{A}}_{p_2}^{\dagger,u}\hat{A}_{0}^{\dagger,u}\hat{\tilde{B}}_{0}^{\dagger,v}\hat{B}_{p_2}^{\dagger,v}
    \\&+
    \Phi_2 c_3 \sqrt{1-\eta_2}
    \sqrt{\eta_3}
    \hat{\tilde{A}}_{p_2}^{\dagger,u}
    \hat{\tilde{A}}_{0}^{\dagger,u}
    \hat{B}_{0}^{\dagger,v}
    +
    \Phi_2 \Phi_3 \sqrt{1-\eta_2}\sqrt{\eta_3}\hat{\tilde{A}}_{p_2}^{\dagger,u}\hat{\tilde{A}}_{0}^{\dagger,u}\hat{B}_{0}^{\dagger,v}\hat{B}_{p_2}^{\dagger,v}
    \end{aligned}
\end{equation}
\\

\emph{Conditional fidelity} --- 
We can now calculate the fidelities ~\eqref{eq:GHZ_non_mixing_fidelity} and \eqref{eq:Cluster_non_mixing_fidelity}, with the photons emitted during the excitation sequence playing a role of the environment. Since the photons emitted in different time bins are orthogonal, the only off-diagonal terms in~\eqref{eq:GHZ_non_mixing_fidelity} and \eqref{eq:Cluster_non_mixing_fidelity} which survive the trace operation are
\begin{equation}
\begin{aligned}
\label{eq:off_diagonal_excitation_errors_appendix}
    &\mathrm{Tr}_{\mathrm{pulse}}
    \int_0^{\infty}
    \mathrm{d} t
    \bra{\emptyset}
    \hat{a}_{u}(t)
    \hat{P}_{n > 0}
    \hat{\tilde{Q}}_{u}^{\dagger}
    \ket{\emptyset}
    \bra{\emptyset}
    (\hat{P}_{n > 0}
    \hat{\tilde{Q}}_{v}^{\dagger})^{\dagger}
    \hat{a}_{v}^{\dagger}(t)
    \ket{\emptyset}
    \\&=
    \int_0^{\infty}
    \mathrm{d} t
    \bra{\emptyset}
    \hat{a}_{u}(t)
    \hat{P}_{n > 0}
    (
    c_1c_3\sqrt{\eta_3}\hat{B}^{\dagger,v}_0
    +
    c_0c_2\sqrt{\eta_2}\hat{A}^{\dagger,u}_0
    )
    \ket{\emptyset}
    \bra{\emptyset}
    (
    c_1c_3\sqrt{\eta_3}\hat{B}^{\dagger,u}_0
    +
    c_0c_2\sqrt{\eta_2}\hat{A}^{\dagger,v}_0
    )^{\dagger}
    \hat{a}_{v}^{\dagger}(t)
    \ket{\emptyset}
    =
    \eta_2
    |c_0 c_2|^2.
    \end{aligned}
\end{equation}
The diagonal terms will contain contribution from all terms that include $\hat{A}_0^{\dagger}$ in~\eqref{eq:Q_operator_fully_filtered_appendix}. Since none of these terms interfere, the diagonal terms is given by the sum of the corresponding coefficients square,
\begin{equation}
\begin{aligned}
\label{eq:diagonal_excitation_errors_appendix}
    &\mathrm{Tr}_{\mathrm{pulse}}
    \int_0^{\infty}
    \mathrm{d} t
    \bra{\emptyset}
    \hat{a}_{u}(t)
    \hat{P}_{n > 0}
    \hat{\tilde{Q}}_{u}^{\dagger}
    \ket{\emptyset}
    \bra{\emptyset}
    (\hat{P}_{n > 0}
    \hat{\tilde{Q}}_{u}^{\dagger})^{\dagger}
    \hat{a}_{u}^{\dagger}(t)
    \ket{\emptyset}
    \\&=
    \eta_2
    (
    |c_0 c_2|^2
    +
    |c_0 \Phi_2|^2
    +
    |c_2 \Phi_0|^2
    +
    |\Phi_0 \Phi_2|^2
    )
    +
    \eta_2(1-\eta_3)
    (|c_3 c_2|^2
    +
    |c_3 \Phi_2|^2
    +
    |c_2 \Phi_3|^2
    +
    |\Phi_2 \Phi_3|^2).
    \end{aligned}
\end{equation}

Finally, for the calculation of the detection probability  all terms in~\eqref{eq:Q_operator_fully_filtered_appendix} contribute and none of the terms interfere, therefore the success probability is given by the sum of the square of all coefficients  in~\eqref{eq:Q_operator_fully_filtered_appendix} and becomes~\eqref{eq:success_prob_second_order}. 
\\

\emph{Calculations of the wavefunction coefficients} --- 
In order to obtain an expression for the conditional fidelity affected by imperfect excitation process, we need to calculate all the coefficients in~\eqref{eq:Q_operator}. Following a wave-function ansatz method of Ref.~\cite{das2019wavefunction} and after some algebra, the coupled differential equations for the first-order coefficients become
\begin{equation}
    \begin{aligned}
        \frac{\partial}{\partial \tau} c_e(\tau)  &= i\frac{\tilde{\Omega}}{2}c_g(\tau) - \Big{(}\frac{1}{2} + i\tilde{\Delta} \Big{)}c_e(\tau)
        \\
        \frac{\partial}{\partial \tau} c_g(\tau)  &= i\frac{\tilde{\Omega}}{2}c_e(\tau)
        \\
        c_e(0) &= 0
        \\
        c_g(0) &= 1,
    \end{aligned}
\end{equation}
while the second-order coefficients are governed by
\begin{equation}
    \begin{aligned}
        \phi_g(\tau,\tau_e) &= ie^{-i\tilde{\Delta}\tau}c_e(\tau_e)\theta(\tau_e - \tau)
        \\
        \phi_e(\tau,\tau_e) &= 0
    \end{aligned}
\end{equation}
for $\tau < \tau_e$ and
\begin{equation}
    \begin{aligned}
        \frac{\partial}{\partial \tau} \phi_e(\tau,\tau_e)  &= i\frac{\tilde{\Omega}}{2}\phi_g(\tau,\tau_e) - \Big{(}\frac{1}{2} + i\tilde{\Delta} \Big{)}\phi_e(\tau,\tau_e)
        \\
        \frac{\partial}{\partial \tau} \phi_g(\tau,\tau_e)  &= i\frac{\tilde{\Omega}}{2}\phi_e(\tau,\tau_e)
        \\
        \phi_g(\tau_e,\tau_e) &= ie^{-i\tilde{\Delta}\tau_e}c_e(\tau_e)
        \\
        \phi_e(\tau_e,\tau_e) &= 0
    \end{aligned}
\end{equation}
for $\tau > \tau_e$, respectively. The dimensionless units used above are $\tau = \gamma t$, $\tilde{\Delta} = \Delta/\gamma$, and $\tilde{\Omega} = \Omega/\gamma$. The equations above have been analytically solved in  first-order perturbation theory for a square-shaped pulse in Ref.~\cite{Pol.Msc.2019}. Adjusting the square pulse to the optimal duration $T_{\mathrm{opt,sq}} = \sqrt{3}\pi/\Delta$, which ensures that a $2\pi$ Rabi oscillation has been performed on the off-resonant transition, while the the resonant transition performs a $\pi$ rotation, the wavefunction coefficients read
\begin{equation}
\begin{aligned}
|c_3|^2 &= 0
\quad\text{}\quad 
|c_0|^2 = 1
\quad\text{}\quad 
|c_1|^2 = \frac{\sqrt{3}\pi}{2\tilde{\Delta}}
\quad\text{}\quad 
|c_2|^2 = 1- \frac{\sqrt{3}\pi}{2\tilde{\Delta}}
\\
|\Phi_2|^2 &= \frac{\sqrt{3}\pi}{8\tilde{\Delta}}
\Big{(}1-\frac{\sqrt{3}\pi}{2\tilde{\Delta}}\Big{)}
\quad
|\Phi_1|^2 = \frac{3\sqrt{3}\pi}{8\tilde{\Delta}}
-
\frac{3\pi^2}{2\tilde{\Delta}^2}
\Big{(}\frac{3}{8}-\frac{1}{\pi^2}\Big{)}
\\
|\Phi_3|^2 &= 
\frac{3}{16}
\Big{(}
\frac{\sqrt{3}\pi}{8\tilde{\Delta}}-\frac{3\pi^2}{16\tilde{\Delta}^2}\Big{)}
\quad\text{}\quad 
|\Phi_0|^2 = 
\frac{13\sqrt{3}\pi}{128\tilde{\Delta}}
\Big{(}
1 - \frac{\sqrt{3}\pi}{2\tilde{\Delta}}
\Big{)}.
\end{aligned}
\end{equation}
Inserting these coefficients into Eq.~\eqref{eq:fidelity_with_second_order_emission} and expanding up to the first order in $\gamma/\Delta$, we arrive at an expression for the fidelities of both the GHZ and the cluster states
\begin{equation}
\begin{aligned}
    \tilde{\mathcal{F}}_{\mathrm{exc,sq}} = 1 - \frac{N\gamma\sqrt{3}\pi}{256\Delta}
    \Big{(}
    29
    +
    3
    \Big{[}
    1 
    +
    \frac{\xi_3}{\xi_2}
    \Big{]}
    \Big{)}.
\end{aligned}
\end{equation}
With perfect frequency filters $\xi_2 = 1$, $\xi_3 = 0$, this turns into
\begin{equation}
\begin{aligned}
\label{eq:excitation_no_finters_1st_order}
    \tilde{\mathcal{F}}_{\mathrm{exc,sq}} = 1 - N\frac{\gamma\sqrt{3}\pi}{8\Delta}.
\end{aligned}
\end{equation}

\section{Fidelities of the states with branching errors}\label{sec:Appendix_C}

\subsection{GHZ state with branching error}\label{subsec:appendix_D_GHZ}

Substituting~\eqref{eq:GHZ_branching_N_operator} and its Hermitian conjugate into Eq.~\eqref{eq:exp_fidelity} yields
    \begin{equation}
    \begin{aligned}
    \nonumber
    &\mathcal{F}^{(N)}_{}
    =
    \mathrm{Tr}_{\mathrm{photons}}
    \Big{\{}
    \int_0^{\infty}
    \mathrm{d} t_N
    ..
    \int_0^{\infty}
    \mathrm{d} t_1
    \bra{\emptyset}
    \bra{\Psi_0}
    \hat{o}_1(t_1)
    ..
    \hat{o}_N(t_N)
    \hat{O}^{\dagger}_N
    ..
    \hat{O}^{\dagger}_1
    \ket{\Psi_0}
    \ket{\emptyset}
    \bra{\emptyset}
    \bra{\Psi_0}
    \hat{O}_1
    ..
    \hat{O}_N
    \hat{o}^{\dagger}_N(t_N)
    ..
    \hat{o}^{\dagger}_1(t_1)
    \ket{\Psi_0}
    \ket{\emptyset}
    \Big{\}}
    \\&=
    \frac{1}{4}
    \mathrm{Tr}_{\mathrm{photons}}
    \Big{\{}
    \int_0^{\infty}\mathrm{d} t_1
    ..
    \int_0^{\infty}\mathrm{d} t_N
    \bra{\emptyset}
    \Big{[}
    \prod_{j=1}^{N}
    \hat{a}_{e,j}(t_{j})
    \Big{(}
    \sqrt{p_{\parallel}}
    \hat{A}_{e,j}^{\dagger}
    +
    \sqrt{p_{\perp}^{\prime}p_{\perp}}
    \hat{A}_{e^{\prime}j}^{\dagger}
    \hat{L}_j^{\dagger}
    \Big{)}
    +
    {p_{\parallel}}^{N/2}
    \prod_{j=1}^{N}
    \hat{a}_{l,j}(t_{j})
    \hat{A}_{l,j}^{\dagger}
    +
    \sqrt{p_{\perp}^{\prime}p_{\parallel}}
    {p_{\parallel}}^{(N-1)/2}
    \prod_{j=1}^{N}
    \hat{a}_{l,j}(t_{j})
    \hat{A}_{l,j}^{\dagger}
    \hat{E}_j^{\dagger}
    \Big{]}
    \ket{\emptyset}
    \\&
    \bra{\emptyset}
    \Big{[}
    \prod_{i=1}^{N}
    \hat{a}_{e,i}(t_i)
    \Big{(}
    \sqrt{p_{\parallel}}
    \hat{A}_{e,i}^{\dagger}
    +
    \sqrt{p_{\perp}^{\prime}p_{\perp}}
    \hat{A}_{e^{\prime}i}^{\dagger}
    \hat{L}_i^{\dagger}
    \Big{)}
    +
    {p_{\parallel}}^{N/2}
    \prod_{i=1}^{N}
    \hat{a}_{l,i}(t_i)
    \hat{A}_{l,i}^{\dagger}
    +
    \sqrt{p_{\perp}^{\prime}p_{\parallel}}
    {p_{\parallel}}^{(N-1)/2}
    \prod_{i=1}^{N}
    \hat{a}_{l,i}(t_i)
    \hat{A}_{l,i}^{\dagger}
    \hat{E}_i^{\dagger}
    \Big{]}^{\dagger}
    \ket{\emptyset}
    \Big{\}}
    \\&=
    \frac{1}{4}
    \mathrm{Tr}_{\mathrm{photons}}
    \Big{\{}
    \prod_{j=1}^{N}
    \int_0^{\infty}\mathrm{d} t_j
    \bra{\emptyset}
    \hat{a}_{e,j}(t_{j})
    \Big{(}
    \sqrt{p_{\parallel}}
    \hat{A}_{e,j}^{\dagger}
    +
    \sqrt{p_{\perp}^{\prime}p_{\perp}}
    \hat{A}_{e^{\prime}j}^{\dagger}
    \hat{L_j^{\dagger}}
    \Big{)}
    \ket{\emptyset}
    \bra{\emptyset}
    \Big{(}
    \sqrt{p_{\parallel}}
    \hat{A}_{e,j}
    +
    \sqrt{p_{\perp}^{\prime}p_{\perp}}
    \hat{A}_{e^{\prime}j}
    \hat{L}_j
    \Big{)}
    \hat{a}_{e,j}^{\dagger}(t_j)
    \ket{\emptyset}
    \\&+
    \frac{p_{\parallel}^N}{4}
    \prod_{j=1}^{N}
    \int_0^{\infty}\mathrm{d} t_j
    \bra{\emptyset}
    \hat{a}_{l,j}(t_j)
    \hat{A}_{l,j}^{\dagger}
    \ket{\emptyset}
    \bra{\emptyset}
    \hat{A}_{l,j}
    \hat{a}_{l,j}^{\dagger}(t_j)
    \ket{\emptyset}
    +
    \frac{p_{\parallel}^N
    p_{\perp}^{\prime}}{4}
    \prod_{j=1}^{N}
    \int_0^{\infty}\mathrm{d} t_j
    \bra{\emptyset}
    \hat{a}_{l,j}(t_j)
    \hat{A}_{l,j}^{\dagger}
    \ket{\emptyset}
    \bra{\emptyset}
    \hat{A}_{l,j}
    \hat{a}_{l,j}^{\dagger}(t_j)
    \ket{\emptyset}
    \\&+
    \frac{p_{\parallel}^{N/2}}{4}
    \Big{[}
    \prod_{j=1}^{N}
    \int_0^{\infty}\mathrm{d} t_j
    \bra{\emptyset}
    \hat{a}_{e,j}(t_{j})
    \Big{(}
    \sqrt{p_{\parallel}}
    \hat{A}_{e,j}^{\dagger}
    +
    \sqrt{p_{\perp}^{\prime}p_{\perp}}
    \hat{A}_{e^{\prime}j}^{\dagger}
    \hat{L_j^{\dagger}}
    \Big{)}
    \ket{\emptyset}
    \bra{\emptyset}
    \hat{A}_{l,j}
    \hat{a}_{l,j}^{\dagger}(t_j)
    \ket{\emptyset}
    +
    h.c.
    \Big{]}
    \Big{\}}
    \\&=
    \frac{\Big{(}p_{\parallel} + {p_{\perp}^{\prime} p_{\perp}}\Big{)}^N
    +
    p_{\parallel}^N
    \Big{(}
    3
    +
    p_{\perp}^{\prime}\Big{)}}
    {4}.
    \end{aligned}
\end{equation}

\subsection{Cluster state with branching error}\label{subsec:appendix_D_Cluster}

A single round of the cluster-state preparation protocol in the presence of imperfect branching updates the state according to Eq.~\eqref{eq:cluster_single_round_branching}
\begin{equation}
    \begin{aligned}
    \label{eq:cluster_single_round_branching_app}
    \hat{O}_j^{\dagger}
    &=
    \ket{+}\bra{1}
    \Big{(}
    \sqrt{p_{\parallel}}
    \hat{A}^{\dagger}_{e,j}
    +
    \sqrt{p_{\perp}^{\prime}p_{\perp}}
    \hat{A}^{\dagger}_{e^{\prime},j}
    \hat{L}_j^{\dagger}
    \Big{)}
    +
    \ket{-}\bra{0}
    \sqrt{p_{\parallel}}
    \hat{A}^{\dagger}_{l,j}
    +
    \ket{-}\bra{1}
    \sqrt{p_{\perp}^{\prime}p_{\parallel}}
    \hat{A}^{\dagger}_{l,j}
    \hat{E}_j^{\dagger}\delta_{j,1}.
    \end{aligned}
\end{equation}
As indicated by the Kronecker-delta, the last term in the equation above only gives a non-zero contribution to the fidelity when the first photon is generated. To calculate the fidelity, we first ignore this term. The operator $\hat{o}(t_1)\hat{O}^{\dagger}$ in Eq.~\eqref{eq:exp_fidelity} corresponding to as single-photon state reads
\begin{equation}
    \begin{aligned}
        \hat{o}_1(t_1)\hat{O}_1^{\dagger}
        &=
        \sqrt{p_{\parallel}}
        \Big{(}
        \ket{1}
        \bra{1}
        \hat{a}_{e,1}(t)
        \hat{A}^{\dagger}_{e,1}
        +
        \ket{0}
        \bra{0}
        \hat{a}_{l,1}(t)
        \hat{A}^{\dagger}_{l,1}
        \Big{)}
        +
        \sqrt{p_{\perp}p_{\perp}'}
        \ket{1}
        \bra{1}
        \hat{a}_{e,1}(t)
        \hat{A}^{\dagger}_{e',1}
        \hat{L}^{\dagger}
        \end{aligned}
\end{equation}
The corresponding fidelity reads
\begin{equation}
\begin{aligned}
    \mathcal{F}^{(1)} &=
    \int_0^{\infty}
    \mathrm{d} t_1
    \textrm{Tr}_{\textrm{photons}}
    \Big{\{}
    \bra{\Psi_0}
    \hat{o}_1(t_1)
    \hat{O}_1^{\dagger}
    \ket{\Psi_0}
    \bra{\emptyset}
    \bra{\emptyset}
    \bra{\Psi_0}
    \hat{O}_1 
    \hat{o}^{\dagger}_1(t_1)
    \bra{\Psi_0}
    \Big{\}}
    =
    p_{\parallel}
    +
    \frac{p_{\perp}^{\prime}p_{\perp}}{4}.
\end{aligned}
\end{equation}
Analogously, for two photons, 
\begin{equation}
    \begin{aligned}
        \hat{o}_{1}(t_{1})
        \hat{o}_2(t_2)
        \hat{O}_1^{\dagger}
        \hat{O}_2^{\dagger}
        &=
        \frac{{p_{\parallel}}}{2}
        \Big{(}
        \ket{1}
        \bra{1}
        \hat{a}_{e,1}(t)
        \hat{A}^{\dagger}_{e,1}
        +
        \ket{0}
        \bra{0}
        \hat{a}_{l,{1}}(t)
        \hat{A}^{\dagger}_{l,1}
        \Big{)}
        \Big{(}
        \hat{a}_{e,2}(t)
        \hat{A}^{\dagger}_{e,2}
        +
        \hat{a}_{l,2}(t)
        \hat{A}^{\dagger}_{l,2}
        \Big{)}
        \\&+
        \frac{\sqrt{p_{\perp}p_{\perp}'p_{\parallel}}}{2}
        \Big{[}
        \Big{(}
        \ket{1}
        \bra{1}
        \hat{a}_{e,1}(t)
        \hat{A}^{\dagger}_{e,1}
        +
        \ket{0}
        \bra{0}
        \hat{a}_{l,{1}}(t)
        \hat{A}^{\dagger}_{l,1}
        \Big{)}
        \hat{a}_{e',2}(t)
        \hat{A}^{\dagger}_{e',2}
        \hat{L}_2^{\dagger}
        \\&+
        \ket{1}
        \bra{1}
        \hat{a}_{e,1}(t)
        \hat{A}^{\dagger}_{e,1}
        \hat{L}^{\dagger}_{1}
        \Big{(}
        \hat{a}_{e,2}(t)
        \hat{A}^{\dagger}_{e,2}
        +
        \hat{a}_{l,2}(t)
        \hat{A}^{\dagger}_{l,2}
        \Big{)}
        \Big{]}
        \\&+
        \frac{p_{\perp}p_{\perp}'}{2}
        \ket{1}\bra{1}
        \hat{a}_{e,2}(t)
        \hat{A}^{\dagger}_{e,2}
        \hat{L}^{\dagger}_{2}
        \hat{a}_{e,1}(t)
        \hat{A}^{\dagger}_{e,1}
        \hat{L}^{\dagger}_{1}
        \end{aligned},
\end{equation}
and the corresponding fidelity is
\begin{equation}
\begin{aligned}
    {\mathcal{F}}^{(2)} &=
    \int_0^{\infty}
    \mathrm{d} t_2
    \textrm{Tr}_{\textrm{photons}}
    \Big{\{}
    \bra{\Psi_0}
    \hat{o}_1(t_{1})
    \hat{o}_2(t_2) 
    \hat{O}_2^{\dagger}
    \hat{O}_{1}^{\dagger}
    \ket{\Psi_0}
    \ket{\emptyset}
    \bra{\emptyset}
    \bra{\Psi_0}
    \hat{O}_{1}
    \hat{O}_2 
    \hat{o}^{\dagger}_2(t_2)
    \hat{o}^{\dagger}_{1}(t_{1})
    \ket{\Psi_0}
    \Big{\}}
    =
    \Big{(}
    p_{\parallel}
    +
    \frac{p_{\perp}^{\prime}p_{\perp}}{4}\Big{)}^2.
\end{aligned}
\end{equation}
Repeating the same procedure $N$ times, the $N$-photon unconditional fidelity reads
\begin{equation}
\begin{aligned}
    {\mathcal{F}}^{(N)} 
    &=
    \Big{(}
    p_{\parallel}
    +
    \frac{p_{\perp}^{\prime}p_{\perp}}{4}\Big{)}^N.
\end{aligned}
\end{equation}
Finally, we multiply by the probability for the first photon to be emitted via the process described by the last term in Eq.~\eqref{eq:cluster_single_round_branching_app}, which yields the total unconditional fidelity of the cluster state in the presence of imperfect branching~\eqref{eq:fidelity_of_cluster_branching_unconditional},
\begin{equation}
    \begin{aligned}
    \label{eq:fidelity_of_cluster_branching_unconditional_app}
    \mathcal{F}^{(N)}[\mathrm{Cl}]
    =
    \Big{(}
    p_{\parallel}
    +
    \frac{p_{\perp}p_{\perp}'}{4}
    \Big{)}^{N-1}
    \Big{(}
    p_{\parallel}
    +
    \frac{p_{\perp}p_{\perp}'}{4}
    +
    \frac{p_{\parallel}p_{\perp}'}{4}
    \Big{)}.
    \end{aligned}
\end{equation}

\section{Branching error decomposition}\label{sec:appendix_E} 
As discussed in the main text dephasing and  and excitation errors can be seen as single qubit errors affecting only a single photon. In this appendix we analyse the nature of the branching errors, which flips the spin state. Since the spin acts as an entangler between subsequently emitted photons,
errors in spin operation could potentially lead to delocalization of such errors between many photonic qubits. Below we show
that this is not the case, and imperfect branching only introduces effective two-qubit errors between two subsequently emitted
photons.

Consider all possible processes arising from branching errors. After a single round of the protocol, the state takes the form of Eq.~\eqref{eq:single_round_branching}, where all but two terms occur from various unwanted decay processes. For the realistic experimental parameters used in Fig.~\ref{fig:8}, only one of these processes occurs with a non-vanishing probability $P = {p_{\perp}^{\prime}p_{\parallel}}$ and is described by an operator 
\begin{equation}\label{eq:Otilde}
    \hat{\tilde{O}}^{\dagger} = \hat{R}\ket{1}\bra{1} \hat{A}^{\dagger}_{l,j} \hat{E}_j^{\dagger}.
\end{equation}
Here we choose $\hat{R} = \hat{H}$ that produces the cluster state, but the same analysis applies to the GHZ state. The remaining coefficients in Eq.~\eqref{eq:single_round_branching} are at least two orders of magnitude smaller and such event therefore almost never appear in real situation. Ignoring these terms, we can represent a single repetition of the protocol with 
\begin{equation}
    \hat{\rho}
    \rightarrow
    p_{\parallel}
    \hat{O}^{\dagger}_{\mathrm{id}}
    \hat{\rho}
    \hat{O}_{\mathrm{id}}
    +
    p_{\perp}^{\prime}p_{\parallel}
    \hat{\tilde{O}}^{\dagger}
    \hat{\rho}
    \hat{\tilde{O}}
    =
    p_{\parallel}
    \hat{O}^{\dagger}_{\mathrm{id}}
    \hat{\rho}
    \hat{O}_{\mathrm{id}}
    +
    p_{\perp}^{\prime}p_{\parallel}
    \hat{\mathcal{E}}^{\dagger}
    \hat{O}^{\dagger}_{\mathrm{id}}
    \hat{\rho}
    \hat{O}_{\mathrm{id}}
    \hat{\mathcal{E}}
\end{equation}
where $\hat{O}^{\dagger}_{\mathrm{id}}$ is the ideal operation of the protocol, $\hat{\tilde{O}}^{\dagger}$ is defined in Eq.~\eqref{eq:Otilde}, and $\hat\rho$ is the systems density matrix. Hence, to prove that imperfect branching introduces at most two-photon errors, one needs to show that $\hat{\mathcal{E}}^{\dagger}$ is a two-qubit operator.

Consider two photons emitted in the ideal protocol, transforming the state according to
\begin{equation}
    \hat{O}^{\dagger}_{\mathrm{id},j+1}
    \hat{O}^{\dagger}_{\mathrm{id},j}
    =
    \ket{+}
    a^{\dagger}_{e,j+1}
    \Big{(}
    a^{\dagger}_{e,j}
    \bra{1}
    -
    a^{\dagger}_{l,j}
    \bra{0}
    \Big{)}
    +
    \ket{-}
    a^{\dagger}_{l,j+1}
    \Big{(}
    a^{\dagger}_{e,j}
    \bra{1}
    +
    a^{\dagger}_{l,j}
    \bra{0}
    \Big{)}.
\end{equation}
On the other hand, a photon emitted through the wrong process of Eq.~\eqref{eq:Otilde} followed by a correctly emitted photon corresponds to
\begin{equation}
    \hat{O}^{\dagger}_{\mathrm{id},j+1}
    \hat{\tilde{O}}^{\dagger}_{j}
    =
    \Big{(}
    \ket{+}
    \bra{1}
    a^{\dagger}_{e,j+1}
    +
    \ket{-}
    \bra{0}
    a^{\dagger}_{l,j+1}
    \Big{)}
    \ket{-}
    \bra{1}
    a^{\dagger}_{l,j}
    =
    \ket{-}\bra{1}
    a^{\dagger}_{l,j+1}
    a^{\dagger}_{l,j}
    -
    \ket{+}\bra{1}
    a^{\dagger}_{e,j+1}
    a^{\dagger}_{l,j}.
\end{equation}
Comparing the two equations above, one can notice that  
\begin{equation}
    \hat{O}^{\dagger}_{\mathrm{id},j+1}
    \hat{\tilde{O}}^{\dagger}_{j}
    =
    -a^{\dagger}_{l,j}a_{e,j}
    \hat{Z}_{j+1}
    \hat{O}^{\dagger}_{\mathrm{id},j+1}
    \hat{O}^{\dagger}_{\mathrm{id},j}.
\end{equation}
Therefore, the spin-flip error of Eq.~\eqref{eq:Otilde} effectively applies a two-photon error between subsequently emitted photons $j$ and $j+1$,
\begin{equation}
    \hat{\mathcal{E}}^{\dagger}
    =
     -a^{\dagger}_{l,j}a_{e,j}\hat{Z}_{j+1}.
\end{equation} 
This error simultaneously flip the phase of photon $j$ and exchanges the logical state of photon $j+1$. Any subsequent operations commute with this error term, since they only affect later photons. The final state of the protocol can thus be understand as an error affecting an ideal state after its preparation.


As was noted earlier, for realistic experimental parameters used in this paper, any error other that~\eqref{eq:Otilde} occurs with a vanishinlgy small probability. Hence, incorrect branching operation always applies a two qubit error $\hat{\mathcal{E}}$, which according to Fig.~\eqref{fig:7} is 0.3\%.

\twocolumngrid
\bibliographystyle{apsrev4-1}
\bibliography{reflist}
\end{document}